\pdfoutput=1
\documentclass[conference]{IEEEtran}
\usepackage[T1]{fontenc}
\usepackage{graphicx}
\usepackage{booktabs}
\usepackage{amsmath}
\usepackage{amssymb}
\usepackage{multirow}
\usepackage{url}
\usepackage{tikz}
\usetikzlibrary{arrows.meta,positioning,fit,backgrounds,calc}
\usepackage{algorithm}
\usepackage{algpseudocode}
\usepackage[hidelinks]{hyperref}

\newcommand{\lab}[1]{\textsf{#1}}

\newcommand{\ClientType}{c7i.4xlarge}
\newcommand{\ServerType}{i4i.4xlarge}
\newcommand{\Region}{ap-south-2}
\newcommand{\ClientCPU}{Intel Xeon Platinum 8488C}
\newcommand{\ClientVCPU}{16}
\newcommand{\ClientRAMGiB}{31}
\newcommand{\PythonVersion}{3.12.3}
\newcommand{\CryptographyVersion}{50.0.1}
\newcommand{\MinioPyVersion}{7.2.20}
\newcommand{\LiboqsVersion}{0.16.0}
\newcommand{\LiboqsPyVersion}{0.16.0.1}
\newcommand{\MinioRelease}{RELEASE.2025-10-15T17-29-55Z}
\newcommand{\LinkGbit}{12.4}

\newcommand{\RTTms}{0.33}
\newcommand{\FioReadMiB}{2{,}044}
\newcommand{\FioWriteMiB}{882}

\newcommand{\TotalRuns}{1{,}970}
\newcommand{\TotalRequestsM}{101.3}

\newcommand{\HsGetRone}{1.486}

\newcommand{\HsPutPone}{0.045}
\newcommand{\HsGetPone}{0.047}
\newcommand{\HsRoneOverPoneGet}{32}

\newcommand{\NetCeilingMiB}{1{,}477}
\newcommand{\NetCeilingPct}{100}
\newcommand{\PutSmallCeiling}{4{,}700}

\newcommand{\PutSmallClientPct}{49}
\newcommand{\PutSmallServerPct}{72}
\newcommand{\PutSmallNetPct}{21}

\newcommand{\StallSlowMin}{12}
\newcommand{\StallSlowMax}{24}
\newcommand{\PutMidCeilingMiB}{1{,}310}

\newcommand{\GetSmallPoneClientPct}{93}

\newcommand{\RoneGetThrMin}{0.38}
\newcommand{\RoneGetThrMax}{0.59}
\newcommand{\HoneCells}{44}
\newcommand{\HoneEquiv}{38}
\newcommand{\HoneInconc}{6}
\newcommand{\HoneDiff}{0}

\newcommand{\HoneSatGetPct}{3.9}

\newcommand{\KeShareSmallPut}{1.41}
\newcommand{\KeShareLargePut}{0.012}
\newcommand{\KeShareSmallGet}{2.19}
\newcommand{\KeShareLargeGet}{0.010}
\newcommand{\PoneSmallPutMs}{3.18}

\newcommand{\HthreeSmallOnePct}{73}

\newcommand{\HthreeMidOnePct}{34}

\newcommand{\HfourBtwoMin}{$-$0.045}
\newcommand{\HfourBtwoMax}{0.048}
\newcommand{\HfourRoneGetMin}{1.37}
\newcommand{\HfourRoneGetMax}{1.63}

\newcommand{\MemPeakSmallSat}{2{,}060}
\newcommand{\MemModeSpreadPct}{2.5}
\newcommand{\ServerGrowthOverlap}{4}
\newcommand{\ServerGrowthWorkloads}{4}
\newcommand{\ServerGrowthOneRun}{756}
\newcommand{\ServerGrowthOtherRuns}{20}
\newcommand{\MemGrowthLargeOne}{20}
\newcommand{\NoneSaveOneMin}{30}
\newcommand{\NoneSaveOneMax}{66}

\newcommand{\NoneGetGainMin}{1.4}
\newcommand{\NoneGetGainMax}{5.1}

\newcommand{\QaKemRatioLow}{5.5}

\newcommand{\QaRateLow}{1{,}500}

\newcommand{\LocTotal}{8{,}006}
\newcommand{\TestCount}{290}

\newcommand{\RevRuns}{1{,}165}

\newcommand{\RevSettleMedian}{3.0}
\newcommand{\RevLinkGbit}{12.4}
\newcommand{\RevRtt}{0.32}
\newcommand{\RevDiskRead}{1{,}970}
\newcommand{\RevDiskWrite}{850}
\newcommand{\ArmRuns}{330}

\newcommand{\ArmLinkGbit}{14.9}
\newcommand{\ArmRtt}{0.16}
\newcommand{\ArmDiskRead}{2{,}010}
\newcommand{\ArmDiskWrite}{870}
\newcommand{\RevRegion}{ap-southeast-2}
\newcommand{\RevClientCPU}{Intel Xeon Platinum 8488C}

\newcommand{\ArmClientCPU}{AWS Graviton3}

\newcommand{\RevAllRuns}{1{,}495}
\newcommand{\RevAllRequests}{97.6}
\newcommand{\RevBatchERuns}{300}
\newcommand{\RevBatchEKRuns}{80}
\newcommand{\RevBatchFRuns}{380}
\newcommand{\RevBatchGRuns}{195}
\newcommand{\RevBatchHRuns}{210}
\newcommand{\RevBatchIRuns}{210}
\newcommand{\RevBatchIKRuns}{60}
\newcommand{\RevBatchIOneRuns}{60}
\newcommand{\RevPrimaryEq}{6}
\newcommand{\RevPrimaryInc}{2}

\newcommand{\RevPrimaryEqTwoPct}{5}

\newcommand{\RevPrimaryNoStallFlips}{0}
\newcommand{\RevPrimaryStalledPairs}{1}
\newcommand{\RevHotRatio}{0.960}
\newcommand{\RevHotDeficit}{4.0}
\newcommand{\RevHotCI}{0.959--0.962}
\newcommand{\RevNoisyRatio}{1.000}

\newcommand{\RevKneeRatio}{1.017}

\newcommand{\RevKneeP}{0.041}
\newcommand{\RevKneePHolm}{0.083}
\newcommand{\RevKneeSd}{8}
\newcommand{\RevLargeRatioLo}{0.998}
\newcommand{\RevLargeRatioHi}{1.001}
\newcommand{\RevXwingHotDeficit}{9.6}
\newcommand{\RevXwingMidDeficit}{4.5}
\newcommand{\RevXwingEq}{6}

\newcommand{\RevXwingVsMlkemHotDeficit}{5.8}
\newcommand{\ArmPrimaryEq}{6}
\newcommand{\ArmPrimaryInc}{2}

\newcommand{\ArmPrimaryEqTwoPct}{4}
\newcommand{\ArmPrimaryEqTenPct}{7}
\newcommand{\ArmPrimaryNoStallFlips}{1}
\newcommand{\ArmKneeNoStallPHolm}{0.076}
\newcommand{\ArmPrimaryStalledPairs}{3}

\newcommand{\ArmHotDeficit}{4.9}
\newcommand{\ArmHotCI}{0.950--0.953}

\newcommand{\ArmNoisySd}{15}

\newcommand{\ArmLargeRatioLo}{0.992}
\newcommand{\ArmLargeRatioHi}{1.001}
\newcommand{\ArmXwingHotDeficit}{10.4}
\newcommand{\ArmXwingMidDeficit}{4.9}
\newcommand{\ArmXwingEq}{4}

\newcommand{\ArmXwingVsMlkemHotDeficit}{5.9}
\newcommand{\RevStudyHybridEq}{8}
\newcommand{\RevStudyHybridLo}{0.969}
\newcommand{\RevRsaGetLo}{0.42}
\newcommand{\RevRsaGetHi}{0.46}
\newcommand{\RevKeEcdhPut}{0.089}

\newcommand{\RevKeEcdhGet}{0.049}

\newcommand{\RevKeMLKEMPut}{0.045}

\newcommand{\RevKeMLKEMGet}{0.047}

\newcommand{\ArmKeMLKEMPut}{0.058}

\newcommand{\ArmKeMLKEMGet}{0.063}

\newcommand{\RevKeStudyHybridPut}{0.133}

\newcommand{\RevKeStudyHybridGet}{0.097}

\newcommand{\RevKeXwingPut}{0.142}

\newcommand{\RevKeXwingGet}{0.133}
\newcommand{\RevKeShareXwingGet}{5.8}

\newcommand{\ArmKeShareXwingGet}{8.2}
\newcommand{\RevXwingOverMlkemPut}{3.1}
\newcommand{\RevXwingOverMlkemGet}{2.8}
\newcommand{\RevKeSumGet}{0.095}
\newcommand{\RevKeShareXwingMidPut}{1.5}
\newcommand{\RevKeShareXwingLarge}{0.10}

\newcommand{\RevMatrixXwingLo}{0.91}
\newcommand{\RevMatrixXwingHi}{1.02}
\newcommand{\RevMatrixXwingWithin}{15}
\newcommand{\RevMatrixCells}{18}
\newcommand{\RevMatrixTests}{162}
\newcommand{\RevMatrixRawSig}{66}
\newcommand{\RevMatrixFdrSig}{50}
\newcommand{\RevMidPutStudyHybridDeficit}{10}
\newcommand{\RevMidPutXwingDeficit}{6}
\newcommand{\RevMatrixXwingCpuHi}{21}
\newcommand{\RevHybridRssDiffMedian}{4}
\newcommand{\RevHybridRssDiffMedianPct}{0.4}
\newcommand{\RevHybridRssDiffMaxPct}{5}

\newcommand{\RevMetaMLKEM}{2{,}034}
\newcommand{\RevMetaStudyHybrid}{2{,}085}

\newcommand{\RevMetaExtraMLKEM}{1{,}412}

\newcommand{\RevMetaXwingOverMLKEM}{40}
\newcommand{\RevMetaIntrinsic}{1{,}056}

\newcommand{\RevMetaEncodingOverhead}{356}
\newcommand{\RevMetaBaseSixtyFour}{1{,}452}
\newcommand{\RevMetaNameBytes}{4}
\newcommand{\RevMetaIntrinsicBillionTB}{1.06}

\newcommand{\RevMetaBillionTB}{1.41}
\newcommand{\RevMetaBillionTiB}{1.28}

\newcommand{\RevSessCpuSavingLo}{2.1}
\newcommand{\RevSessCpuSavingHi}{3.7}
\newcommand{\RevSessThroughputLo}{0.966}
\newcommand{\RevSessThroughputHi}{1.062}

\newcommand{\RevSessConfigs}{9}
\newcommand{\RevSessBoneCpu}{1.7}
\newcommand{\RevSessShortKem}{0.0038}
\newcommand{\RevSessShortObjects}{248}

\newcommand{\RevSessLightObjects}{94}

\newcommand{\RevSessLightHundredObjects}{94}
\newcommand{\RevSessFullKem}{0.0011}

\newcommand{\RevSessGetGain}{3.3}
\newcommand{\RevSessGetGainLo}{3.1}
\newcommand{\RevSessGetGainHi}{3.5}

\newcommand{\RevQSmallRhoKneeHi}{0.81}

\newcommand{\RevQMidRhoKneeLo}{0.68}

\newcommand{\RevProfFutexTime}{45}
\newcommand{\RevProfNamespaceTime}{35}
\newcommand{\RevProfNamespaceCalls}{65}
\newcommand{\RevProfWriteTime}{5}
\newcommand{\RevProfSyncTime}{1.1}
\newcommand{\RevProfSyncCalls}{0.5}
\newcommand{\RevProfStraceCycles}{14}
\newcommand{\RevProfDiskAwait}{0.20}
\newcommand{\RevProfDiskAwaitPninetyfive}{0.58}
\newcommand{\RevProfDiskUtil}{76}
\newcommand{\RevProfDiskQueue}{3.0}
\newcommand{\RevProfDiskSeconds}{292}
\newcommand{\RevProfDiskWriteMib}{283}
\newcommand{\RevProfCpuUser}{68}
\newcommand{\RevProfCpuKernel}{31}
\newcommand{\RevProfCpuKernelLocks}{4}

\newcommand{\RevOrderShiftMax}{1.1}

\newcommand{\RevCvQuietMax}{2.2}
\newcommand{\RevCvNoisyLo}{4}
\newcommand{\RevCvNoisyHi}{9}
\newcommand{\RevPowerNoisy}{3}

\newcommand{\RevPowerNoisyNeeded}{50}
\newcommand{\RevStalls}{8}

\newcommand{\ArmStalls}{3}

\newcommand{\RevNoisyCpuCorr}{0.91}
\newcommand{\RevNoisyRuns}{40}

\newcommand{\RevPutStageManifestDelta}{16.2}

\newcommand{\RevGetStageManifestDelta}{6.4}
\newcommand{\RevGetStageKdfDelta}{4.6}

\newcommand{\ArmPutStageManifestDelta}{16.4}

\newcommand{\ArmGetStageManifestDelta}{5.5}
\newcommand{\ArmGetStageKdfDelta}{4.6}
\newcommand{\RevModelErrHot}{0.9}
\newcommand{\RevModelErrHotRsa}{1.7}
\newcommand{\RevModelErrMidLo}{1.8}
\newcommand{\RevModelErrMidHi}{3.0}
\newcommand{\RevModelErrMidRsa}{3.5}

\newcommand{\RevHotCpuDelta}{79}
\newcommand{\RevHotCpuPerRequest}{1.71}

\newcommand{\ArmHotCpuDelta}{109}

\newcommand{\RevCpuPrimaryExcl}{3}
\newcommand{\ArmCpuPrimaryExcl}{6}
\newcommand{\RevBootAgree}{15}
\newcommand{\RevBootTotal}{16}
\newcommand{\RevOpenRuns}{185}

\newcommand{\RevMarginUtil}{0.71}

\newcommand{\AudRuns}{480}

\newcommand{\AudPadRatio}{0.993}
\newcommand{\AudPadCI}{0.991--0.995}
\newcommand{\AudPadCost}{0.7}
\newcommand{\AudRestRatio}{0.967}
\newcommand{\AudRestCI}{0.965--0.969}
\newcommand{\AudRestCost}{3.3}
\newcommand{\AudJFullRatio}{0.960}

\newcommand{\AudPadShare}{17}
\newcommand{\AudPadShareLo}{14}
\newcommand{\AudPadShareHi}{21}
\newcommand{\AudPadVerdict}{does not explain}
\newcommand{\AudPredicted}{0.956}
\newcommand{\AudPredictionErrorJ}{0.4}

\newcommand{\AudPadStageGet}{6.0}
\newcommand{\AudMlkemStageGet}{5.6}
\newcommand{\AudKdfOverPadGet}{4.3}
\newcommand{\AudJRuns}{120}
\newcommand{\AudKKneeRatio}{0.994}
\newcommand{\AudKKneeCI}{0.984--1.004}

\newcommand{\AudKSatRatio}{0.996}
\newcommand{\AudKSatCI}{0.979--1.013}

\newcommand{\AudKRuns}{200}
\newcommand{\AudKStalls}{6}
\newcommand{\AudBindVerdict}{explains}
\newcommand{\AudLFullRatio}{0.963}
\newcommand{\AudLFullCI}{0.960--0.967}

\newcommand{\AudPredictionErrorL}{0.7}
\newcommand{\AudPredictionInsideL}{outside}
\newcommand{\AudBindReleasedRatio}{0.999}
\newcommand{\AudBindReleasedCI}{0.997--1.002}

\newcommand{\AudBindReleasedRemoved}{99}
\newcommand{\AudBindReleasedRemovedLo}{92}
\newcommand{\AudBindReleasedRemovedHi}{105}

\newcommand{\AudBindReleasedKeOne}{0.028}
\newcommand{\AudBindHeldRatio}{1.002}
\newcommand{\AudBindHeldCI}{0.997--1.006}

\newcommand{\AudBindHeldOverReleased}{1.002}
\newcommand{\AudBindHeldOverReleasedCI}{0.999--1.006}

\newcommand{\AudBindBindingKeOne}{0.047}
\newcommand{\AudBindEcdhKeOne}{0.050}
\newcommand{\AudLRuns}{120}
\newcommand{\AudAllocDefaultLo}{0.12}
\newcommand{\AudAllocDefaultHi}{0.38}
\newcommand{\AudAllocDefaultSpread}{3.2}
\newcommand{\AudAllocDefaultThrSpread}{10}
\newcommand{\AudAllocFixedLo}{0.29}
\newcommand{\AudAllocFixedHi}{0.32}
\newcommand{\AudAllocFixedSpread}{1.1}
\newcommand{\AudAllocFixedThrSpread}{3}
\newcommand{\AudAllocRuns}{40}
\newcommand{\AudNRuns}{90}

\newcommand{\AudNGo}{1.26.2}

\newcommand{\AudNOneCpuUs}{343}

\newcommand{\AudNOneMlkemThr}{0.905}
\newcommand{\AudNOneMlkemThrCI}{0.900--0.910}
\newcommand{\AudNOneMlkemVerdict}{deficit}
\newcommand{\AudNOneMlkemDeficit}{9.5}
\newcommand{\AudNOneMlkemCpu}{1.105}

\newcommand{\AudNOneXwingThr}{0.826}
\newcommand{\AudNOneXwingThrCI}{0.821--0.831}

\newcommand{\AudNOneXwingDeficit}{17.4}
\newcommand{\AudNOneXwingCpu}{1.211}

\newcommand{\AudNMidCores}{9.75}

\newcommand{\AudNMidMlkemDeficit}{5.7}

\newcommand{\AudNMidXwingDeficit}{7.7}

\newcommand{\AudNHotCores}{11.50}

\newcommand{\AudNHotXwingDeficit}{1.1}

\newcommand{\AudNOneMlkemCpuDelta}{36}
\newcommand{\AudNOneXwingCpuDelta}{72}
\newcommand{\AudNMicroMlkem}{50.6}
\newcommand{\AudNMicroEcdh}{33.4}
\newcommand{\AudNMicroXwing}{85.1}
\newcommand{\AudNMicroMlkemExtra}{17}
\newcommand{\AudNMicroXwingExtra}{52}
\newcommand{\AudSweepCells}{102}
\newcommand{\AudSweepDeficits}{7}
\newcommand{\AudSweepAdvantages}{2}
\newcommand{\AudSweepDeficitMinCores}{14.2}
\newcommand{\AudSweepDeficitMinWorkers}{128}
\newcommand{\AudSweepDeficitMaxSizeKiB}{64}
\newcommand{\AudSweepDeficitOps}{GET}

\newcommand{\AudSweepAdvantageOps}{PUT}
\newcommand{\AudSweepAdvantageHi}{1.06}
\newcommand{\AudSweepAdvantageLo}{1.04}
\newcommand{\AudSweepBusyNoDeficit}{4}

\newcommand{\finding}[2]{\par\smallskip\noindent\fbox{\parbox{\dimexpr\columnwidth-2\fboxsep-2\fboxrule\relax}{\small\textbf{Finding~#1.} #2}}\par\smallskip}

\begin{document}

\title{Benchmarking Post-Quantum Key Encapsulation in S3-Compatible Object Storage under Client Concurrency}

\newcommand{\cbitauthor}[2]{\begin{minipage}[t]{0.3\textwidth}\centering
{\normalsize #1}\\[2pt]
{\footnotesize\textit{Dept. of Information Technology}\\
\textbf{\textit{Chaitanya Bharathi Institute of Technology}}\\
Hyderabad, India\\
#2}\end{minipage}}
\author{%
\cbitauthor{Revanth Reddy Lingala}{revanthreddy0906@gmail.com}\hfill
\cbitauthor{K. Gangadhara Rao}{kgangadhar\_it@cbit.ac.in}\hfill
\cbitauthor{Padigela Srujan}{padigela.srujan2007@gmail.com}\\[10pt]
\hspace*{\fill}\cbitauthor{Saarthak Manocha}{saarthakmanocha@gmail.com}\hspace{0.05\textwidth}%
\cbitauthor{Vainavi Venkat}{vainavivenkat06@gmail.com}\hspace*{\fill}}

\maketitle

\begin{abstract}
Data encrypted today with classical public-key algorithms could be decrypted once large quantum
computers exist. Cloud object stores keep data for years, so the client-side encryption that
protects it must move to post-quantum algorithms such as ML-KEM, which adds a key-encapsulation
step and a larger ciphertext to every stored object. So far, the cost of ML-KEM has been
measured for the algorithm on its own and for securing network connections, but not for
encrypting each object in a storage system that serves many clients at once. It has therefore
been unclear whether ML-KEM slows such a system, under which conditions, and why. We address
this problem by measuring per-object ML-KEM against classical key exchange on an S3-compatible
object store, with two client implementations, two processor types and up to 500 concurrent
workers, using pre-registered experiments and statistical equivalence tests. ML-KEM reduced
throughput only when the client's processor was at or near full load, with one unexplained
exception, and then by \RevHotDeficit{} to \ArmHotDeficit\,\% in our main client. Controlled
experiments on one processor type traced most of that loss to the way the software library was
called, and a second, native client showed that the size of the loss depends on the
implementation. The lasting cost is storage: \RevMetaIntrinsic{} more bytes per object than with
classical key exchange. The
results cover two clients, one single-node store and one cloud provider.
\end{abstract}


\section{Introduction}
Cloud object stores keep data for years, and much of that data is encrypted under keys
established with RSA or elliptic-curve Diffie--Hellman. Shor's algorithm would allow a
sufficiently large quantum computer to recover those keys, so an adversary who records
ciphertext today can decrypt it once such a machine exists, a threat usually called
``harvest now, decrypt later''. For data at rest the exposure lasts as long as the data is
retained. In August 2024 NIST published FIPS~203, standardizing the module-lattice
key-encapsulation mechanism ML-KEM~\cite{fips203}.

With client-side envelope encryption, as implemented by the Amazon S3 Encryption Client and by
Azure Blob Storage client libraries~\cite{s3ec,azurecse}, every object is encrypted under its own
data key, and that key is protected for the recipient and stored with the object.

A public-key KEM per object is needed only in some deployments. If writer and reader share a symmetric
key-encryption key, wrapping the data key with AES is already quantum-safe and no KEM is needed;
that is our baseline B1. A public-key mechanism is required when the writer must not be able to
decrypt, or does not share a secret with the reader: write-only producers such as log shippers,
backup agents and sensors that hold only a public key; objects encrypted to several recipients
or to an escrow key; and data handed across organizational boundaries. These deployments use RSA
or elliptic-curve key establishment today, and they are the ones that must migrate. For them,
ML-KEM places a key establishment and its 1{,}088-byte ciphertext on every PUT and every GET
rather than once per connection.

What is known falls into three groups of measurements. Primitive-level studies report the cost of key generation, encapsulation
and decapsulation and the sizes of keys and ciphertexts on several processor
families~\cite{liboqs,rodriguez2025,demir2025,lopez2025}. Studies of post-quantum TLS and KEMTLS
measure connection establishment under latency, loss and bandwidth limits
\cite{paquin2020,sikeridis2020,sosnowski2023,giron2023}. Storage studies measure the S3 interface
without cryptography~\cite{gadban2021} and client-side encryption without post-quantum key
establishment~\cite{henziger2019,ozeri2023}.

What is not known is the cost on the data path of an object store. None of these measures ML-KEM where an object store would use it: once
per object, on a data path whose throughput is set jointly by client CPU, network, metadata
handling and the storage server, with many clients at once. A back-of-envelope estimate already
says that tens of microseconds of KEM against milliseconds of storage round trip should be
invisible. What that estimate cannot say is whether it holds under concurrency, in which regime
it fails, by how much, and whether the cause is the KEM computation or the larger ciphertext it
carries through the data path (Section~\ref{sec:related}, Table~\ref{tab:gap}).

The thesis of this paper is the following. In the configurations we measured, replacing X25519 by per-object ML-KEM lowered
throughput only where the client was at or near CPU saturation, with one unexplained exception
in the native reader. The size of the loss there depended on the client implementation: it
ranged from no measurable
loss (\AudBindReleasedRatio{} of X25519's throughput, the Python client calling liboqs through
one foreign call) through \RevHotDeficit--\ArmHotDeficit\,\% (the same client through the
standard binding) to \AudNOneMlkemDeficit\,\% (a native Go reader on one core). The cost that
every implementation pays is storage: a larger ciphertext with every object. The observations
that do not fit this picture are collected in Section~\ref{sec:anomalies}.

The contributions are the following.
\begin{itemize}
\item[C1] The saturation-regime result, with a method to apply it: measurements
of ML-KEM-768 against X25519 over object size, 1--500 concurrent workers, two client
architectures and two client implementations, with the independently restarted run as the
experimental unit and pre-registered equivalence tests. We use a CPU accounting identity
(Section~\ref{sec:cpumethod}) as a method: under client saturation the throughput ratio of two
modes is the inverse ratio of their client CPU per request, which turns a migration question
into a CPU measurement and a saturation check
(Sections~\ref{sec:primary}--\ref{sec:native}).
\item[C2] A binding artifact, and what it implies for library authors: two
pre-registered controls on the x86 Python client show that its client-bound deficit is produced
by a library binding that makes four foreign calls per decapsulation, and for
\AudPadShare\,\% by the larger manifest (Section~\ref{sec:mechanism}).
\item[C3] The storage cost: the per-object metadata cost of ML-KEM and of two
hybrids, separated into ciphertext and encoding, scaled to fleet sizes, and set against the
2\,KB limit that Amazon S3 places on user-defined object metadata
(Section~\ref{sec:metadata}).
\end{itemize}
Secondary results are the measured trade-off of bounded session amortization, which saves a few
percent of client CPU and changes throughput only on a client-bound workload
(Section~\ref{sec:tradeoff}); the attribution of the store's small-object write ceiling to its
metadata path and not to its disk (Section~\ref{sec:bottleneck}); and the observation that
RSA-4096 is the one mode that moves the bottleneck in the Python client. The instrumented client, the harness, the native reader and the
analysis pipeline are an artifact that supports these results and are not claimed as a
contribution; neither is the architecture or commit protocol, which is not a new cryptographic
construction (Table~\ref{tab:claims}). The open-loop capacity observations in the supplementary
material describe the store, not post-quantum cryptography, and their main lesson is
known~\cite{schroeder2006}.

The scope of the study is limited as follows. We do not evaluate the security of ML-KEM, side channels, or transport security,
which is held constant. All performance statements concern one Python client stack and one small native reader, one
single-node MinIO deployment and one cloud provider (Section~\ref{sec:validity}). The native
reader covers the GET path of three modes only. Section~\ref{sec:pending} lists the experiments
that remain.

This paper measures what per-object ML-KEM-768 costs against RSA-4096, X25519 and
symmetric-only baselines on the same data path; how that cost depends on object size,
concurrency and a hybrid construction, and what causes it; and what bounded sessions trade
for the key-establishment work they save.

\section{Background and Related Work}\label{sec:background}\label{sec:related}
\subsection{Post-Quantum KEMs, Hybrids and HPKE}\label{sec:kems}
A key-encapsulation mechanism (KEM) consists of key generation, encapsulation, which takes the
recipient's public key and outputs a ciphertext and a shared secret, and decapsulation, which
recovers the secret with the private key. ML-KEM, derived from CRYSTALS-Kyber, relies on the
hardness of module learning with errors. ML-KEM-768 targets NIST security category~3 with a
1{,}184-byte encapsulation key, a 1{,}088-byte ciphertext and a 32-byte shared
secret~\cite{fips203}. Decapsulation uses \emph{implicit rejection}: a modified ciphertext yields
a pseudorandom secret instead of an error, so tampering is detected by the authenticated
encryption that follows.

A hybrid KEM combines a classical and a post-quantum component so that the result stays secure
while either remains unbroken. The combiner matters: X-Wing~\cite{barbosa2024,xwing} combines
X25519 with ML-KEM-768 as
$\mathrm{SHA3\mbox{-}256}(ss_M\|ss_X\|ct_X\|pk_X\|\mathit{label})$ and derives both private keys
from one 32-byte seed; its security proof depends on these concrete choices and does not carry
over to other ways of mixing two secrets.

Hybrid Public Key Encryption (HPKE, RFC~9180)~\cite{rfc9180} is the standard construction for
encrypting to a public key: a KEM produces a shared secret, a key schedule derives an AEAD key
and base nonce from it and from application-supplied context, and messages are sealed under
sequence-numbered nonces. An IETF draft adds ML-KEM-768 and an X25519+ML-KEM-768 hybrid as HPKE
KEMs~\cite{hpkepq}. Per-object envelope encryption has the structure of HPKE's single-recipient
base mode.
HPKE~\cite{rfc9180} and its post-quantum extension~\cite{hpkepq} specify how to encrypt to an
ML-KEM or hybrid public key but contain no measurements; the file-encryption tool age adopted
such recipients~\cite{age2025}. Our envelope predates our reading of that draft and is not
HPKE-conformant; Section~\ref{sec:hpke} maps one onto the other and states what transfers. For
hybrids, the security of the result depends on the combiner. X-Wing~\cite{barbosa2024,xwing}
comes with a proof for its concrete combiner, and the general theory of KEM combiners states the
conditions under which a combination stays secure if one component fails
\cite{giacon2018,bindel2019}.
Our mode P2 concatenates two shared secrets into HKDF and has not been checked against those
conditions, which is why it carries no security claim and serves only as a performance
reference (Table~\ref{tab:claims}).

In envelope encryption a per-object data key encrypts the object, and key establishment protects
that key. We derive data keys with HKDF-SHA256~\cite{rfc5869} and encrypt with AES-256-GCM, whose
security requires that a nonce is never reused under one key~\cite{sp80038d}. The classical
alternatives are key transport with RSA-OAEP~\cite{rfc8017} and ephemeral-static X25519
\cite{rfc7748}. Object stores expose buckets and keys through HTTP operations, chiefly PUT and
GET of whole objects; the S3 interface is the de facto standard~\cite{gadban2021}. We use
MinIO~\cite{minio}, compiled from its final community release (\MinioRelease). Because that
community edition is no longer developed, our store-specific results (ceilings, knee) should be
read as those of one representative single-node S3 server, not as current product behaviour.

The closest analogue of a per-object KEM with bounded reuse is secure messaging. Signal's PQXDH
adds a post-quantum KEM to initial key agreement~\cite{pqxdh2023}, and Apple's PQ3 applies one to
initial key establishment and to a periodic rekeying step as well~\cite{pq32024}. There the unit
of amortization is a session or a rekeying interval, and the question of how many messages one
KEM output may protect is the same one our session mode poses for objects
(Section~\ref{sec:tradeoff}). We are not aware of throughput measurements of these protocols
that would transfer to an object store.

\subsection{Related Measurements}\label{sec:measurements}
Table~\ref{tab:gap} positions this study. The entries describe the scope of each work, not a
shortcoming: a TLS study is not incomplete for omitting S3, but its results cannot establish the
cost of key establishment in an object-store data path.

\begin{table*}[t]
\caption{Scope of related work. \checkmark{} = covered; -- = outside the study's scope or not reported. P = peer-reviewed, A = preprint (arXiv).}
\label{tab:gap}
\centering\footnotesize
\setlength{\tabcolsep}{4pt}
\begin{tabular}{llp{4.6cm}cccccc}
\toprule
Study & Type & Setting & \shortstack{PQ\\KEM} & \shortstack{Hybrid\\KEM} & \shortstack{Object\\store} & \shortstack{Per-object key\\establishment} & \shortstack{Concurrency\\sweep} & \shortstack{Queueing\\analysis} \\
\midrule
Paquin et al.~\cite{paquin2020} & P & TLS 1.3 under emulated latency and loss & \checkmark & \checkmark & -- & -- & -- & -- \\
Sikeridis et al.~\cite{sikeridis2020} & P & Post-quantum authentication in TLS 1.3 & -- & -- & -- & -- & -- & -- \\
Sosnowski et al.~\cite{sosnowski2023} & P & TLS 1.3 handshakes, black- and white-box & \checkmark & \checkmark & -- & -- & -- & -- \\
Giron et al.~\cite{giron2023} & P & Hybrid KEMTLS, simulated and real networks & \checkmark & \checkmark & -- & -- & -- & -- \\
Barbosa et al.~\cite{barbosa2024} & P & X-Wing construction and security proof & \checkmark & \checkmark & -- & -- & -- & -- \\
Rodriguez-Alvarez~\cite{rodriguez2025} & A & ML-KEM against RSA and ECC, primitives & \checkmark & -- & -- & -- & -- & -- \\
Demir et al.~\cite{demir2025} & A & Kyber and Dilithium primitives & \checkmark & -- & -- & -- & -- & -- \\
Lopez et al.~\cite{lopez2025} & A & PQC on resource-constrained devices & \checkmark & -- & -- & -- & -- & -- \\
Gadban and Kunkel~\cite{gadban2021} & P & S3 interface for HPC workloads & -- & -- & \checkmark & -- & \checkmark & -- \\
Henziger and Carlsson~\cite{henziger2019} & P & Client-side encryption and delta encoding & -- & -- & -- & -- & -- & -- \\
Prikutse et al.~\cite{prikutse2026} & A & Hybrid PQ model for multi-file cloud storage & \checkmark & \checkmark & -- & -- & -- & -- \\
Junaid~\cite{junaid2026} & A & Hybrid PQ client-side encryption, self-hosted keys & \checkmark & \checkmark & -- & -- & -- & -- \\
\textbf{This work} & & Client-side envelope encryption on S3 (MinIO) & \checkmark & \checkmark & \checkmark & \checkmark & \checkmark & \checkmark \\
\bottomrule
\end{tabular}
\end{table*}

Existing measurements stop at two boundaries. At the primitive level, the Kyber design
paper~\cite{bos2018}, the standard~\cite{fips203}, library benchmarks~\cite{liboqs} and recent
comparative preprints~\cite{rodriguez2025,demir2025,lopez2025} give the cost and sizes of the
three KEM operations on several processor families; they establish that platform and
implementation matter, which is why we repeat our primary comparison on a second architecture
and test a second implementation, but they contain no storage, concurrency or end-to-end request
latency. At the transport level, studies of post-quantum TLS and KEMTLS
\cite{paquin2020,sikeridis2020,sosnowski2023,giron2023,schwabe2020} measure connection
establishment under latency, loss and bandwidth limits and decompose the handshake. They are the
closest systems precedents, but their amortization boundary differs: a handshake protects a
connection, whereas in client-side object encryption key establishment and its ciphertext recur
for each object.

Storage studies measure the S3 interface without cryptography, and client-side encryption
without post-quantum key establishment. Gadban and Kunkel~\cite{gadban2021} attribute losses on
the S3 path to layers of the client stack, a separation our harness applies to cryptographic
stages; Henziger and Carlsson~\cite{henziger2019} and Ozeri et al.~\cite{ozeri2023} concern
where client-side encryption sits and what it costs synchronization. Two 2026 preprints
approach post-quantum cloud storage from the design side: Prikutse et
al.~\cite{prikutse2026} combine AES-256-GCM, X25519+ML-KEM-768 and ML-DSA-65 and amortize
signatures over a batch of files, and Junaid~\cite{junaid2026} proposes a client-side hybrid
architecture with self-hosted key management. According to their abstracts, neither evaluates
concurrent S3 PUT/GET or compares with classical per-object key establishment; we have read only
the abstracts. Bounded key reuse is established practice in this setting: Amazon S3 Bucket Keys
derive per-object data keys from a time-limited bucket-level key and reduce key-service requests
by up to 99\,\%~\cite{s3bucketkeys}, and the AWS Encryption SDK caches data keys under age,
message and byte limits~\cite{esdkcache}. Our session mode applies that principle to ML-KEM
roots and quantifies the trade-off.

In October 2026 we ran keyword searches, restricted to arXiv, dblp and Semantic Scholar and on
the open web, combining \emph{ML-KEM}, \emph{Kyber} or \emph{X-Wing} with \emph{object storage},
\emph{S3}, \emph{MinIO}, \emph{client-side encryption}, \emph{envelope encryption} and
\emph{concurrency}; the queries and their outcome are logged in the artifact. This is a keyword
search, not a systematic review. It returned primitive benchmarks, transport studies, vendor
tutorials and the two storage designs above, but, to our knowledge, no measurement of per-object
ML-KEM under concurrent S3 load. The search did not cover IEEE Xplore, the ACM Digital Library,
the IACR ePrint archive or Google Scholar, which is a limitation of that statement.

\subsection{Benchmarking Method}\label{sec:benchmethod}
Three strands of method inform our design. Cloud instances show run-to-run and time-dependent
performance variation large enough to change conclusions drawn from single runs
\cite{leitner2016,uta2020,maricq2018},
which is why our experimental unit is the independently restarted run, compared in randomized
blocks and never across sessions. Reporting rules for performance measurements
\cite{hoefler2015,kalibera2013}
ask for intervals over independent repetitions and a full description of the setup. Schroeder et
al.~\cite{schroeder2006} show that closed- and open-loop load generators produce systematically
different behaviour; our open-loop observations are an instance of that result. For equivalence
we follow the two one-sided tests procedure~\cite{schuirmann1987,lakens2017}.

\section{Design and Threat Model}\label{sec:design}\label{sec:n1design}\label{sec:hpke}
The network and the object store are honest-but-curious for confidentiality and potentially
malicious for integrity: an adversary may read, replace, delete, reorder or replay stored
ciphertexts and manifests. The adversary does not hold the recipient's decapsulation key and may
in the future operate a quantum computer. Availability, traffic analysis, deletion of whole
objects, compromised endpoints and side channels in the client are out of scope.

Writers are not authenticated. The envelope encrypts to a public key and carries no
signature, and neither does HPKE base mode. Integrity therefore means that an object written
by an honest writer cannot be modified, reordered, truncated, extended or spliced without
detection. It does not mean that an object is authentic: anyone who knows the recipient's public
key, including the store, can create a new object under any identity, and it will decrypt
without error. Sender authentication is out of scope.

Rollback is not prevented. Replay is detected only where it crosses an identity: a
manifest or chunk moved to another object or another version fails authentication, because the
object identity and version are bound into the key and into every chunk. Serving an
\emph{older, complete and valid version} of the same object is a different matter. A store that
withholds or deletes the newest manifest makes the reader accept the previous version, and
nothing in the design detects that. Freshness would need state outside the store, such as a
signed version counter or a transparency log; it is out of scope.

A third limit concerns key commitment. AES-GCM is not key-committing: a ciphertext can be constructed that
decrypts validly under more than one key
\cite{dodis2018,len2021}.
We accept this for the single-recipient setting studied here, on the following argument. Since
writers are not authenticated, a malicious writer can already store any plaintext it likes
under a key the recipient will derive; a ciphertext that also decrypts under a second key gives
it no further capability against one recipient. That argument is not a proof and we make no
formal claim (Table~\ref{tab:claims}). It does not extend to several recipients, where one
ciphertext could decrypt to different plaintexts for different recipients, or to
password-derived keys: a multi-recipient design must use a
key-committing construction
\cite{albertini2022}.

Table~\ref{tab:modes} lists the modes. All encrypting modes share the HKDF and chunked AES-GCM
path. They differ in the key-establishment computation \emph{and} in the size of the manifest
that carries its ciphertext (622 to 2{,}085 bytes), which changes parsing, hashing and bytes
transferred. A difference between two modes is therefore the cost of key establishment together
with the handling of its ciphertext; Section~\ref{sec:mechanism} separates the two as far as
the data allows. B0 stores plaintext through the same
framing; B1 derives keys from a pre-shared 32-byte key and is the symmetric-only floor; R1
transports a random secret with RSA-OAEP-4096; B2 uses ephemeral-static X25519; P1 uses
ML-KEM-768. Two hybrids combine X25519 with ML-KEM-768. \emph{P3 is X-Wing} as specified in
\cite{xwing}: we implement its key derivation from a 32-byte seed, its SHA3-256 combiner and its
ciphertext layout, and verify key generation, encapsulation and decapsulation against the
draft's published test vectors. \emph{P2 is a study-specific concatenation hybrid}: it stores the
ML-KEM ciphertext followed by the X25519 ephemeral key and feeds both 32-byte secrets,
concatenated, to HKDF together with the object context. P2 predates our X-Wing implementation; we
keep it as a performance reference and make no security claim for its combiner. N1 amortizes one
ML-KEM encapsulation over a bounded session (described below). A reader must name the
mode it expects, so a planted plaintext object cannot downgrade an encrypted one.

\begin{table}[t]
\caption{Key-establishment modes. Sizes in bytes; the per-object value is what every object's manifest carries.}
\label{tab:modes}
\centering\footnotesize
\setlength{\tabcolsep}{3pt}
\resizebox{\columnwidth}{!}{%
\begin{tabular}{lllrrc}
\toprule
Mode & Key establishment & Hard problem & Public key & Per object & PQ \\
\midrule
B0 & none (plaintext) & -- & -- & -- & -- \\
B1 & pre-shared key & -- & -- & 0 & yes \\
R1 & RSA-OAEP-4096 & factoring & 512 & 512 & no \\
B2 & X25519 & ECDLP & 32 & 32 & no \\
P1 & ML-KEM-768 & MLWE & 1{,}184 & 1{,}088 & yes \\
P2 & X25519+ML-KEM-768, concatenation & both & 1{,}216 & 1{,}120 & yes \\
P3 & X-Wing (X25519+ML-KEM-768) & both & 1{,}216 & 1{,}120 & yes \\
N1 & ML-KEM-768 session & MLWE & 1{,}184 & 1{,}088 & yes \\
\bottomrule
\end{tabular}}
\end{table}

Algorithm~\ref{alg:putget} shows PUT and GET for mode P1. An object is identified by
$I=(\mathit{tenant},\mathit{bucket},\mathit{key},\mathit{version})$, where versions are
time-ordered UUIDv7 values. The context $\mathit{ctx}$ is canonical JSON of $I$, the mode, the
algorithm names, chunk size and count, plaintext size, recipient key identifier and the SHA-256
hash of the key-establishment ciphertext $C$; it is bound into the data key and into every chunk.
Objects are split into 4\,MiB chunks, and chunk $j$ is encrypted with nonce
$p\|\mathrm{be}_{64}(j)$ under a random 32-bit prefix $p$ and associated data
$\mathit{ctx}\|j\|\ell_j\|\mathit{final}_j$, so deletion, reordering, truncation, extension and
cross-object splicing of chunks fail authentication. This is the counter-and-final-flag pattern
of the STREAM construction~\cite{hoang2015}, with the flag carried in the associated data
instead of the nonce; we do not claim STREAM's proof for our variant. Each version is stored under
\texttt{key.pqcs/version/} as a ciphertext stream \texttt{data.enc} and a manifest
\texttt{manifest.json} holding $\mathit{ctx}$, $C$, the salt and the prefix. The manifest is
written last and serves as the commit record: readers only consider versions whose manifest
exists, so an interrupted upload is never visible. One logical PUT is therefore two S3 writes
(the data, in parts above 8\,MiB, then the manifest), and one GET of a known version is two S3
reads (the manifest, then the data). Finding the newest version needs an additional LIST, which
the benchmark avoids by reading versions it knows; a client that always asks for ``the latest''
would pay it.

\begin{algorithm}[t]
\caption{Per-object PUT and GET (mode P1)}
\label{alg:putget}
\footnotesize
\begin{algorithmic}[1]
\Procedure{Put}{$\mathit{key}, M, ek$}
\State $v \gets \textsc{UUIDv7}()$; \ $I \gets (\mathit{tenant},\mathit{bucket},\mathit{key},v)$
\State $(C,S) \gets \mathrm{ML\mbox{-}KEM.Encaps}(ek)$
\State $\mathit{salt} \xleftarrow{\$} \{0,1\}^{256}$; \ $p \xleftarrow{\$} \{0,1\}^{32}$
\State $\mathit{ctx} \gets \textsc{Canon}(I,\text{mode},n,|M|,\mathrm{id}(ek),\mathrm{H}(C),\dots)$
\State $K \gets \mathrm{HKDF}(S,\mathit{salt},\text{``pqc-s3-object-v2''}\|0\|\mathit{ctx})$
\For{$j \gets 0$ \textbf{to} $n-1$} \Comment{4\,MiB chunks}
\State $c_j \gets \mathrm{AES\mbox{-}GCM}_K(p\|\mathrm{be}_{64}(j), M_j, \mathit{ctx}\|j\|\ell_j\|\mathit{final}_j)$
\EndFor
\State upload $c_0\|\dots\|c_{n-1}$ in 8\,MiB parts to \texttt{key.pqcs/$v$/data.enc}
\State commit: write $(\mathit{ctx},C,\mathit{salt},p)$ to \texttt{key.pqcs/$v$/manifest.json}
\EndProcedure
\Procedure{Get}{$\mathit{key}, dk, \text{mode}$}
\State $v \gets$ newest version with a manifest; read and parse it
\State \textbf{reject} unless identity, mode, key id and $\mathrm{H}(C)$ all match
\State $S \gets \mathrm{ML\mbox{-}KEM.Decaps}(dk,C)$; \ $K \gets \mathrm{HKDF}(S,\mathit{salt},\dots)$
\For{each chunk $c_j$ of \texttt{data.enc}}
\State $M_j \gets \mathrm{AES\mbox{-}GCM}^{-1}_K(\dots)$; \ \textbf{reject} on failure
\EndFor
\State \textbf{reject} unless exactly $n$ chunks were read and the last is final
\EndProcedure
\end{algorithmic}
\end{algorithm}

In N1 one encapsulation yields a root $S_r$, and object $i$ of the session uses
\[K_i=\mathrm{HKDF}(S_r,\mathit{salt}_i,\text{``pqc-s3-session-v1''}\|0\|\mathit{ctx}_i),\]
with the session identifier, the object's sequence number and the bound authenticated in
$\mathit{ctx}_i$. Every object still stores the session's ML-KEM ciphertext, so it can be
decrypted alone. A root protects at most $B_{\max}$ objects and lives at most $T_{\max}$ seconds,
whichever is reached first. In a queue-aware variant the object bound of each new session is
\[B_t=\mathrm{clip}\bigl(B_{\min},B_{\max},\mathrm{round}(B_{\min}+(B_{\max}-B_{\min})\hat\rho)\bigr),\]
where $\hat\rho$ is a moving average ($\alpha=0.2$) of the fraction of a client process's worker
threads with a request in flight. Receivers cache decapsulated roots under their own caps and
reject objects whose authenticated sequence number exceeds them.

Our envelope has the structure of HPKE base mode~\cite{rfc9180}: one KEM encapsulation, a key
derivation bound to application context, and AEAD encryption of a sequence of records. It
differs in three ways. (1)~The data key is derived with a single HKDF-SHA256 call whose
\emph{info} is the object context, where HPKE uses a labelled extract-and-expand key schedule
that also outputs a base nonce. (2)~Chunk nonces are a random prefix followed by a counter, where
HPKE computes base nonce XOR sequence number. (3)~The chunk index, length and final-chunk flag
are authenticated in the associated data; HPKE leaves truncation protection to the application.
Mode P1 uses a KEM that~\cite{hpkepq} defines for HPKE (ML-KEM-768). That draft also defines an
X25519+ML-KEM-768 hybrid under the KEM identifier that the X-Wing draft~\cite{xwing} requests,
so P3 corresponds to it; we did not test interoperability. A deployed example of this approach is
the file-encryption tool age, whose version 1.3.0 added HPKE-based hybrid ML-KEM-768 recipients
\cite{age2025}. It shows that a per-file post-quantum KEM is already practice; we found no
performance study of it under concurrent load. We did not
implement or benchmark HPKE. Since an HPKE-based envelope would perform the same KEM operation
once per object, a key schedule of a few hash calls and the same AEAD, we expect the per-object
costs reported here to transfer, but that expectation is untested, and a new design should use
HPKE and not this format.

Table~\ref{tab:claims} separates the kinds of claim this paper makes. The engineering properties
are supported by design and by tests, not by a proof of the protocol as a whole.

\begin{table*}[t]
\caption{Claim taxonomy}
\label{tab:claims}
\centering\footnotesize
\begin{tabular}{p{2.5cm}p{7.2cm}p{6.6cm}}
\toprule
Class & Claimed & Not claimed \\
\midrule
Primitive security & Security inherited from ML-KEM-768, HKDF-SHA256 and AES-256-GCM under their stated assumptions and correct use & Any new proof or analysis of these primitives \\
Hybrid construction & P3 implements X-Wing and passes the draft's test vectors, so it inherits that construction's analysis & Any security property of the P2 concatenation combiner; P2 is a performance reference only \\
Envelope & The structure of HPKE base mode (KEM, context-bound key derivation, AEAD) & HPKE conformance, or any analysis of HPKE carrying over to this format \\
Engineering properties & Identity and version binding, authenticated chunk order and final-chunk flag, commit-last visibility, mode-downgrade rejection and session bounds, each exercised by tamper and stress tests & A formal model or proof of the storage protocol; resistance to side channels \\
Experimental findings & Measured latency, throughput, CPU, memory, metadata and queueing behaviour of the evaluated configuration & Performance on other S3 implementations, distributed deployments, native clients other than the GET-only Go reader, other regions or hardware \\
Out of scope & -- & Availability, traffic analysis, compromised endpoints, deletion, sender authentication, rollback to an older valid version, key commitment of the AEAD, and transport security (held constant) \\
\bottomrule
\end{tabular}
\end{table*}

\section{Implementation and Methodology}\label{sec:method}
\subsection{Implementation}\label{sec:impl}
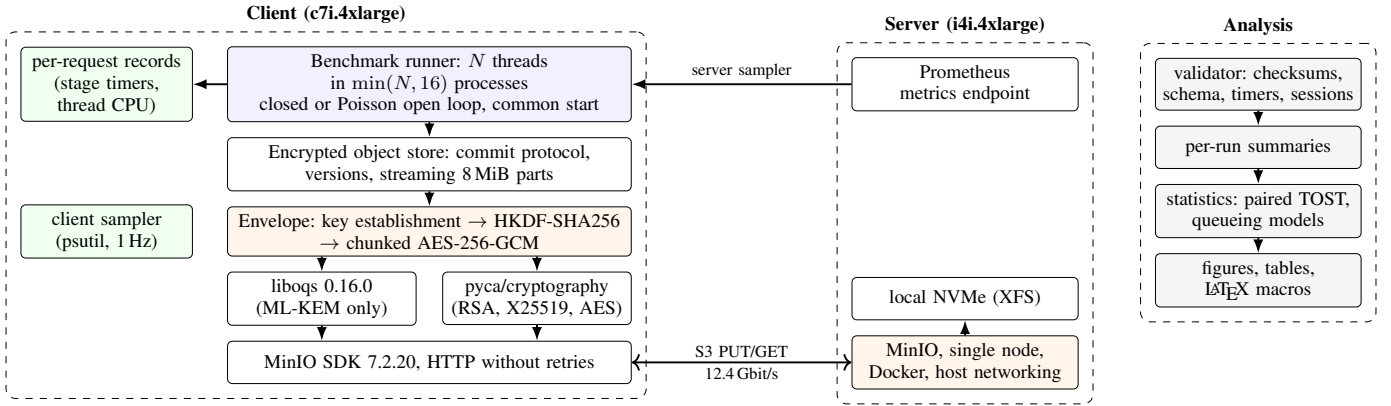
\begin{figure*}[t]
\centering
\resizebox{\textwidth}{!}{%
\begin{tikzpicture}[
  font=\footnotesize,
  box/.style={draw, rounded corners=2pt, align=center, minimum height=0.62cm, inner sep=3pt, fill=white},
  layer/.style={box, text width=5.7cm},
  side/.style={box, text width=2.3cm, fill=green!6},
  lib/.style={box, text width=2.55cm},
  srv/.style={box, text width=3.1cm},
  ana/.style={box, text width=2.8cm, fill=gray!8},
  host/.style={draw, rounded corners=4pt, dashed, inner sep=6pt},
  lbl/.style={font=\scriptsize, fill=white, inner sep=1pt},
  arr/.style={-{Latex[length=2mm]}, thick},
  node distance=0.22cm and 0.5cm]
\node[layer, fill=blue!6] (runner) {Benchmark runner: $N$ threads in $\min(N,16)$ processes\\closed or Poisson open loop, common start};
\node[layer, below=of runner] (store) {Encrypted object store: commit protocol,\\versions, streaming 8\,MiB parts};
\node[layer, below=of store, fill=orange!8] (env) {Envelope: key establishment $\rightarrow$ HKDF-SHA256\\$\rightarrow$ chunked AES-256-GCM};
\node[lib, below=of env.south west, anchor=north west] (oqs) {liboqs \LiboqsVersion\\(ML-KEM only)};
\node[lib, below=of env.south east, anchor=north east] (pyca) {pyca/cryptography\\(RSA, X25519, AES)};
\node[layer, below=of oqs.south west, anchor=north west] (sdk) {MinIO SDK \MinioPyVersion, HTTP without retries};
\node[side, left=of runner] (rec) {per-request records\\(stage timers,\\thread CPU)};
\node[side, left=of env] (csamp) {client sampler\\(psutil, 1\,Hz)};
\node[host, fit=(runner)(sdk)(rec)(csamp), label={[font=\footnotesize\bfseries]above:Client (\ClientType)}] (client) {};
\draw[arr] (runner) -- (store);
\draw[arr] (store) -- (env);
\draw[arr] (env.south -| oqs) -- (oqs);
\draw[arr] (env.south -| pyca) -- (pyca);
\draw[arr] (oqs) -- (oqs |- sdk.north);
\draw[arr] (pyca) -- (pyca |- sdk.north);
\draw[arr] (runner) -- (rec);
\node[srv, anchor=west] (metrics) at ($(client.east |- runner)+(3.0cm,0)$) {Prometheus\\metrics endpoint};
\node[srv, anchor=west, fill=orange!8] (minio) at ($(client.east |- sdk)+(3.0cm,0)$) {MinIO, single node,\\Docker, host networking};
\node[srv, above=of minio] (nvme) {local NVMe (XFS)};
\node[host, fit=(metrics)(minio)(nvme), label={[font=\footnotesize\bfseries]above:Server (\ServerType)}] (server) {};
\draw[arr, <->] (sdk.east) -- (minio.west);
\draw[arr] (metrics.west) -- (runner.east);
\node[lbl, above=1pt] at ($(client.east |- sdk)!0.5!(server.west |- minio)$) {S3 PUT/GET};
\node[lbl, below=1pt] at ($(client.east |- sdk)!0.5!(server.west |- minio)$) {\LinkGbit\,Gbit/s};
\node[lbl, above=1pt] at ($(client.east |- runner)!0.5!(server.west |- metrics)$) {server sampler};
\draw[arr] (minio) -- (nvme);
\node[ana, anchor=west] (val) at ($(server.east |- metrics)+(0.9cm,0)$) {validator: checksums,\\schema, timers, sessions};
\node[ana, below=of val] (sum) {per-run summaries};
\node[ana, below=of sum] (stat) {statistics: paired TOST,\\queueing models};
\node[ana, below=of stat] (paper) {figures, tables,\\\LaTeX{} macros};
\node[host, fit=(val)(paper), label={[font=\footnotesize\bfseries]above:Analysis}] (ana) {};
\draw[arr] (val) -- (sum); \draw[arr] (sum) -- (stat); \draw[arr] (stat) -- (paper);
\end{tikzpicture}}
\caption{System overview (testbed of the exploratory campaign; the later testbeds have the same structure). The client stack (left) issues S3 requests to MinIO (centre); every request produces a record, and client and server resources are sampled once per second. Raw data passes through a validator before any analysis (right), which generates every number in this paper.}
\label{fig:arch}
\end{figure*}

Figure~\ref{fig:arch} shows the system and Table~\ref{tab:components} its components. The client,
harness and analysis are written in Python~\PythonVersion; the project comprises \LocTotal{}
non-blank lines of code, including \TestCount{} automated tests.

\begin{table}[t]
\caption{Components of the implementation (non-blank lines, counted by the analysis pipeline)}
\label{tab:components}
\centering\footnotesize
\setlength{\tabcolsep}{3pt}
\begin{tabular}{lp{4.3cm}r}
\toprule
Component & Responsibility & Lines \\
\midrule
Cryptographic core & ML-KEM wrapper, modes, HKDF, chunked AES-GCM, manifest, sessions & 1{,}000 \\
Storage layer & streaming PUT/GET, commit protocol, versions, garbage collection & 298 \\
Benchmark harness & runner, payloads, samplers, records, validator, micro-benchmarks & 1{,}176 \\
Command-line tool & keys, settings and the \texttt{pqcstore} commands & 412 \\
Testbed automation & AWS provisioning, VM roles, MinIO image & 645 \\
Analysis pipeline & per-run summaries, statistics, paper assets & 3{,}086 \\
Tests & unit, interoperability, tamper, secret-hygiene and integration tests & 1{,}389 \\
\bottomrule
\end{tabular}

\end{table}

ML-KEM-768 comes from liboqs~\LiboqsVersion{} through liboqs-python~\LiboqsPyVersion{}
\cite{liboqs,liboqspython}. We build liboqs from a pinned release with only the ML-KEM parameter
sets enabled and load it from an explicit installation path, so the benchmark cannot silently pick up a different library; a start-up check verifies the library version and that ML-KEM-768 is enabled, and every call checks key and ciphertext sizes.
RSA-OAEP-4096 (SHA-256), X25519, HKDF and AES-256-GCM come from
pyca/cryptography~\CryptographyVersion{}~\cite{pyca}. A registry maps each mode to its
key-establishment and recovery functions, and the envelope module turns the result into a
streaming chunk encryptor or decryptor, so the storage code never holds a whole object. Before
any timed operation every worker process runs a warm-up that loads the libraries, performs a key generation, encapsulation, decapsulation, derivation and encryption round trip, and one full round trip in the mode under test, so that one-off costs never enter a timer.
An interoperability test checks ML-KEM in both directions against an independent implementation
(OpenSSL through pyca), and the manifest parser rejects unknown or missing fields, wrong types, inconsistent sizes and a key-establishment ciphertext that does not match its hash.

X-Wing uses the same liboqs build for its ML-KEM component, through the library's derandomized
key-generation and encapsulation functions, and the same X25519 as mode B2, so a difference
between P3 and P1 or B2 reflects the construction and not a different library. The expanded
decapsulation key is cached per process, which the draft permits; P1 likewise holds its expanded
key.

The storage layer implements the commit protocol of Algorithm~\ref{alg:putget} on the MinIO
Python SDK~\MinioPyVersion. Ciphertext records are passed to the SDK through a file-like adapter
over the encryptor's output, so an object is uploaded in 8\,MiB parts while it is being encrypted;
on GET the stream is decrypted chunk by chunk into a sink. For benchmarking, the HTTP client uses a connection pool sized to the worker threads and performs no automatic retries, so failures are counted rather
than hidden inside latency. Auxiliary functions list committed versions, verify a commit by reading
it back, delete versions and collect uncommitted data left behind by interrupted uploads.

A \emph{run} executes one configuration (mode, operation, object size, concurrency and loop type)
with freshly started worker processes; it is the unit of replication. Concurrency $N$ is realized
as exactly $N$ worker threads spread over $\min(N,16)$ processes, one per vCPU. Each process warms
up, reports readiness, and all of them start at a common wall-clock instant, so that no worker
measures while others are still starting. Each run has a 5\,s warm-up phase and a 20\,s measured
phase; Python's garbage collector is run before and disabled during the timed window, so its
pauses cannot land inside a stage timer. In closed-loop runs each thread issues requests back to
back. In open-loop runs each process draws Poisson arrivals at its share of the target rate, a
request waits in a queue for a free thread, and the wait is recorded separately; a request that
cannot start before the end of the run is recorded as a failure, never dropped. GET runs read a
set of pre-written fixture objects that are created before and deleted after the run, and every
object written by a PUT run is deleted afterwards, outside the timed window.

Every request, including warm-up requests and failures, produces one record with
\texttt{perf\_counter} timers for the five stages of Algorithm~\ref{alg:putget} (key
establishment, key derivation, AES-GCM, manifest handling and storage), its total latency, the
thread CPU time it consumed, bytes on the wire and metadata size, plus identifiers that tie it to
its run, repetition, code commit and configuration. A client sampler records the CPU time and
resident memory of the worker processes and the host's CPU and network counters once per second;
a server sampler reads the same information from MinIO's Prometheus metrics endpoint and, in
the confirmatory campaign, dirty and write-back memory, disk write rate and latency, and I/O wait
from a node-exporter endpoint on the server host.

A validator checks every data set before analysis and refuses it if any check fails: checksums of
all files; a complete schema with correct types for every record; that no request's stage times
exceed its total; that per run the 99th percentile of unaccounted time stays below 2\,\% of the
request time and below 1\,\% in aggregate; and that no N1 session exceeds its bound. The
\TestCount{} automated tests cover the cryptographic core, the commit protocol, tampering (bit flips; deleted, reordered, duplicated, truncated or extended chunks; swapped, replayed or relabelled manifests; mode downgrade), that manifests contain no secret material and private-key files are protected, session bounds under concurrent stress, and
end-to-end runs against a live MinIO.

A single script provisions and operates the cloud testbed: it creates both instances in one
cluster placement group with a firewall that admits SSH only from the operator's address, copies
the committed code, builds MinIO from source into a Docker image, formats the server's local NVMe
disk at boot, runs network (iperf3, ping) and disk (fio) baselines, starts experiments in a
detached session, fetches results and stops or deletes all resources. Before launch, a planner
estimates the run time and cost of each experiment configuration. To bound cost, both instances
stop themselves when idle, and the client stops itself shortly after an experiment completes.

As a second client (Section~\ref{sec:native}) we wrote a deliberately small reader in
Go (\AudNGo): the GET path of modes B2, P1 and P3, with ML-KEM-768, X25519, HKDF, SHA-3 and
AES-GCM from Go's standard library and the MinIO Go SDK. A test opens objects written by the
Python client in all three modes and several sizes and rejects modified ciphertexts and a wrong
expected mode, so the two clients agree on the object format. Because Go's ML-KEM takes keys
as seeds, the recipient keys for this client are generated from seeds on the Python side. Its
benchmark runs goroutines in one process and reports throughput, process CPU and latency; it
has no stage timers.

A summarizer streams the compressed request records of each experiment once and writes one row
per run. A statistics stage computes intervals, paired equivalence tests, bottleneck labels and
the queueing models. A final stage generates every figure, table and number in this paper as
\LaTeX{} macros, so no result is typed by hand.

\subsection{Testbeds and Workloads}\label{sec:campaigns}
The study has an exploratory campaign, a confirmatory campaign and a set of follow-up
experiments, which we never pool.

The \emph{exploratory campaign} (\TotalRuns{} runs, \TotalRequestsM{} million
requests) covers the full matrix with five repetitions per configuration: 1--500 workers,
1\,KiB--500\,MiB objects, RSA and symmetric baselines, sessions and an open loop. Its equivalence
verdicts used a 95\,\% interval inside $\pm$5\,\% and no multiplicity control; we report them as
exploratory.

The \emph{confirmatory campaign} (\RevAllRuns{} runs, \RevAllRequests{} million
requests, none of which failed, including the \RevOpenRuns{} open-loop runs, where every queued
request was eventually served) was pre-registered: hypotheses, primary cells, test, margin,
multiplicity correction, repetitions, stall rule, the training and held-out split for the
queueing model and the session grid were written into the project's versioned specification and
committed before any of its data existed. Table~\ref{tab:batches} maps the purpose labels used
in this paper to the original batch identifiers of the artifact.

The timestamps of the pre-registration show one thing and leave others open. The commit that contains the
pre-registration was pushed to a hosted repository (GitHub) on 10~October 2026 at 16:19~UTC, and
the first run of the campaign started on 11~October at 11:53~UTC. The hosting service's
repository activity log records each push with its time and resulting commit; the authors
cannot edit it, and an export of it is part of the artifact. This establishes one thing: the registered text existed on a third
party's server before the first run. It does not establish what the authors had seen locally
before pushing, it is not a registration with an independent registry, and it covers a
repository that was private while the experiments ran and was made public only at submission, so outsiders could not observe the pushes as they happened. One
registration was pushed after a run of another experiment had begun
(Section~\ref{sec:deviations}, item~4). Analyses that were not pre-specified are listed in
Section~\ref{sec:deviations} and are exploratory.

\begin{table}[t]
\caption{Experiments by purpose label, with the batch identifiers used in the artifact and the pre-registration. Exploratory tables and figure titles generated by the pipeline keep the batch identifiers A to D; the run counts of the exploratory campaign are in Table~\ref{tab:sources}.}
\label{tab:batches}
\centering\footnotesize
\setlength{\tabcolsep}{2.5pt}
\begin{tabular}{lllrp{3.0cm}}
\toprule
Label & Batch & Client & Runs & Purpose \\
\midrule
\lab{Expl-Sweep} & A & x86 & & concurrency sweep \\
\lab{Expl-Size} & B & x86 & & object size, one worker \\
\lab{Expl-OpenLoop} & C & x86 & & open loop, sessions on GET \\
\lab{Expl-SmallPUT} & D & x86 & & small PUT with 40\,s pause \\
\midrule
\lab{Primary-x86} & E, EK & x86 & \RevBatchERuns, \RevBatchEKRuns & primary family; EK is the knee cell, 20 repetitions \\
\lab{Hybrid-Matrix} & F & x86 & \RevBatchFRuns & B2, P1, P2, P3 \\
\lab{Sessions} & G & x86 & \RevBatchGRuns & $B_{\max}\times T_{\max}$ \\
\lab{Open-Loop} & H & x86 & \RevBatchHRuns & closed-loop $\mu(n)$ and open loop, P1 \\
\lab{Profiling} & P & x86 & -- & profiles only; no performance numbers \\
\lab{Primary-ARM} & I, IK & ARM & \RevBatchIRuns, \RevBatchIKRuns & primary family repeated (B2, P1, P3) \\
\lab{OneWorker-ARM} & I1 & ARM & \RevBatchIOneRuns & one-worker cells \\
\midrule
\lab{Control-Manifest} & J & x86 & \AudJRuns & X25519 with a padded manifest \\
\lab{Power-PUT} & K & x86 & \AudKRuns & 64\,KiB PUT cells, 50 repetitions \\
\lab{Control-Binding} & L & x86 & \AudLRuns & ML-KEM through one foreign call \\
\lab{Allocator} & M, MT & x86 & \AudAllocRuns & diagnostic, 1\,MiB PUT \\
\lab{Native-GET} & N & x86 & \AudNRuns & native (Go) reader, GET only \\
\bottomrule
\end{tabular}
\end{table}

\begin{table}[t]
\caption{Exploratory campaign: runs, requests and the results each batch supplies (batch identifiers as in Table~\ref{tab:batches})}
\label{tab:sources}
\centering\footnotesize
\begin{tabular}{lrrp{4.2cm}}
\toprule
Batch & Runs & Requests & Result of record for \\
\midrule
A & 850 & 35.4\,M & concurrency sweep: GET; PUT $\geq$1\,MiB; hybrid P2 \\
B & 360 & 1.0\,M & object size 1\,KiB--500\,MiB; key-establishment time \\
C & 500 & 46.9\,M & open loop; N1 on GET; queue-aware N1 \\
D & 260 & 18.0\,M & PUT $\leq$64\,KiB incl.\ N1 (40\,s pause) \\
\bottomrule
\end{tabular}

\end{table}

\emph{Follow-up experiments} (\lab{Control-Manifest}, \lab{Power-PUT}, \lab{Control-Binding},
\lab{Allocator} and \lab{Native-GET}; \AudRuns{} runs with the Python client and \AudNRuns{}
with a native one) were added after the confirmatory campaign had shown a client-bound deficit
whose cause we could not identify. Each was pre-registered with its own decision rule, pushed to
the hosted repository before its first run, and is analysed separately; none is pooled with
the confirmatory campaign. They ran on the x86 instance types of the confirmatory campaign with a new server
instance.

In every testbed the client and MinIO run on two instances in one cluster placement group; MinIO
is a single node in Docker with host networking on an \ServerType{} instance (16 vCPU, local
NVMe, XFS). The exploratory campaign ran in AWS region \Region{} with a \ClientType{} client
(\ClientCPU, \ClientVCPU{} vCPU, \ClientRAMGiB\,GiB): \LinkGbit\,Gbit/s between the hosts,
\RTTms\,ms round trip, \FioReadMiB/\FioWriteMiB\,MiB/s disk read/write. The confirmatory campaign
ran in region \RevRegion, on a different account: first with the same instance types
(\RevClientCPU; \RevLinkGbit\,Gbit/s, \RevRtt\,ms, \RevDiskRead/\RevDiskWrite\,MiB/s), then with a
c7g.4xlarge client (\ArmClientCPU, 16 vCPU) and a new server of the same type
(\ArmLinkGbit\,Gbit/s, \ArmRtt\,ms, \ArmDiskRead/\ArmDiskWrite\,MiB/s). Because the ARM testbed
has a different server instance and network path, we compare ratios between modes across
architectures, never absolute throughput.

PUT and GET are measured separately, mostly in a closed loop. Each run has a 5\,s warm-up and a
20\,s measured phase. The exploratory campaign measures concurrency 1--500 at 64\,KiB and
1\,MiB (16\,MiB up to 128 workers) and object size 1\,KiB--500\,MiB at one worker. The
confirmatory campaign measures the primary cells defined below; a hybrid matrix of 1\,KiB to
128\,MiB at 1 and 128 workers; sessions at $B_{\max}\in\{1,10,100,1000\}$ and
$T_{\max}\in\{1,10,60\}$\,s; and, for the queueing analysis reported in the supplementary
material, closed-loop throughput at 1--256 workers and Poisson arrivals at 12 rates from
1{,}500 to 5{,}600 requests/s served by 256 threads, for 1\,KiB and 64\,KiB PUT. The
pre-registered queueing protocol (alternate rates as training and held-out points, scored by
error on the stable held-out rates) is stated in full there.

The metrics are defined as follows.
\begin{itemize}
\item \emph{Per-object key-establishment time}: the time a request spends in encapsulation, RSA
unwrapping, Diffie--Hellman or decapsulation, from the stage timers. It is sometimes loosely
called ``handshake overhead''; we avoid that name because no multi-message protocol is
involved.
\item \emph{Throughput}: successful requests, or MiB, per second of the measured window.
\item \emph{CPU}: CPU of the client's benchmark processes and of MinIO during the measured
window, and client thread CPU time per request.
\item \emph{Memory}: peak resident memory and its growth over the run.
\item \emph{Response time} (open loop): queue wait plus service time per request.
\item \emph{Metadata}: bytes of the manifest stored with each object.
\end{itemize}
Above one worker, stage timers include time spent waiting for the Python interpreter lock, so
key-establishment time and stage shares are reported at one worker only.

A run starts fresh worker processes and is the experimental unit. The millions of requests inside
a run sharpen that run's own estimate; they are not independent replicates, and no interval in
this paper is computed over requests. Runs are organized in randomized complete blocks: one
repetition contains every mode of a cell once, in shuffled order, and modes are compared only
within a repetition.

Between runs the harness waits until the server host's dirty plus write-back memory is below
32\,MiB and its disk write rate below 5\,MiB/s for two consecutive seconds, at most 40\,s. In the
exploratory experiments \lab{Expl-Sweep}, \lab{Expl-Size} and \lab{Expl-OpenLoop} this pause had silently fallen back to 5\,s, which slowed some PUT runs
that followed another PUT run by \StallSlowMin--\StallSlowMax\,\%; \lab{Expl-SmallPUT} repeated the affected
cells with a fixed 40\,s pause. In the confirmatory campaign the measured pause had a median of
\RevSettleMedian\,s.

\subsection{Statistical Analysis}\label{sec:tost}\label{sec:cpumethod}\label{sec:deviations}
For a cell and two modes $a$ and $b$, repetition $i$ gives the paired log ratio
$d_i=\log(x_{a,i}/x_{b,i})$ of throughput. We test equivalence with two one-sided tests
(TOST)~\cite{schuirmann1987,dixon2018} against the bounds $[\log 0.95,\log 1.05]$:
\[H_0^{-}:\ \mathrm{E}[d]\le\log 0.95,\qquad H_0^{+}:\ \mathrm{E}[d]\ge\log 1.05,\]
each with a one-sided paired $t$-test at $\alpha=0.05$. The modes are declared \emph{equivalent}
only if both are rejected, which is the same as the 90\,\% confidence interval of the ratio lying
inside $[0.95,1.05]$; the TOST $p$-value is the larger of the two. We also show the 95\,\%
interval for estimation. A cell whose 90\,\% interval lies wholly outside the bounds is
\emph{non-equivalent}; anything else is \emph{inconclusive}, which is not evidence of ``no
difference''.

The margin of five percent is our smallest effect of interest and was fixed in advance, as the
margin the exploratory campaign had already used. Its practical meaning is a provisioning one: a
5\,\% loss of capacity raises utilization by the same factor, so a deployment planned at or
below \RevMarginUtil{} of peak throughput stays under the stability boundary we measure for
this store (supplementary material) and needs no additional capacity. That argument uses a result of this
study and was made after the fact; an operator with a tighter target needs a tighter margin,
which is why we report sensitivity at $\pm$2\,\% and $\pm$10\,\%
\cite{lakens2017}.

The pre-registered hypothesis is the ratio P1/B2 in eight
closed-loop cells chosen before the campaign: 64\,KiB GET at 128 and 500 workers; 64\,KiB PUT at
32 workers (the knee of the throughput curve) and 128; 1\,MiB PUT and GET at 32; and 16\,MiB PUT
and GET at 32. The eight $p$-values are adjusted with Holm's procedure~\cite{holm1979}, which
controls the family-wise error rate. A limitation of this family, recognized after the
campaign: only one cell, 64\,KiB GET at 500 workers, has a saturated client. The 64\,KiB PUT
cells are limited by the server and the four cells at 1\,MiB and above by the network or the
server; at 64\,KiB GET with 128 workers the client is the busiest resource but still has spare
CPU, which a costlier mode simply uses (see the CPU accounting below). In seven of eight cells,
therefore, no key-establishment mode could have changed throughput, and equivalence there
confirms where the bottleneck is and little about ML-KEM. Throughput is also
an insensitive endpoint in those cells; we therefore add client CPU per request as an
exploratory second endpoint (Table~\ref{tab:cpuprimary}). The same family measured on the ARM client is a replication
and is adjusted in the same way. All other comparisons (P3 and P2 against P1 and B2, R1 against
P1, and every cell outside the family) are exploratory and reported with unadjusted intervals.

The number of repetitions follows from a power calculation. From the paired log-ratio standard deviations of the exploratory campaign, a
normal-approximation calculation at $\alpha=0.05/8$ and power 0.8 required at most 7 repetitions
in every primary cell except the knee, which required about 19. We therefore ran 10 repetitions
per cell and 20 at the knee, with no optional stopping.

The tests rest on a distributional assumption. The $t$-tests assume approximately normal mean log ratios
from 10 or 20 pairs. As a check we recompute each primary verdict from a percentile-bootstrap
90\,\% interval of the mean log ratio (20{,}000 resamples).

The stall rule was fixed in advance and is blind to the mode: a closed-loop run is a \emph{stall}
if its server CPU is below 0.8 times and its throughput below 0.9 times the median of its own
cell. The primary analysis uses all runs; excluding stalls is a sensitivity analysis.

The second tool of the analysis is a CPU accounting identity.
Throughput is cores in use divided by CPU per request. If the client CPU is exhausted, two modes
use the same number of cores, and the throughput ratio of mode $a$ to mode $b$ is the inverse
ratio of their client CPU per request,
\[\frac{X_a}{X_b}=\frac{c_b}{c_a}.\]
This is an accounting identity, not an empirical finding and not a model with free parameters:
evaluated on the runs it describes, its error measures only how equal the core counts of the two
modes are. We use it as a method, in two ways. It localizes a throughput difference: if the
identity holds, the difference has no source other than client CPU. And it gives an operator a
procedure: measure CPU per request for the current and the candidate mode on the real client,
check whether the client is saturated, and read off the expected change.

The saturation criterion differs between the two clients. For the native reader the criterion was fixed in advance: a cell
is client-saturated if process CPU is at least 0.95 of the cores made available, in every mode.
For the Python client no criterion was pre-registered, and we do not claim a sharp one. In the
64\,KiB GET cells the post-quantum modes use nearly the same number of cores as X25519 at 500
workers and visibly more at 128 workers (Table~\ref{tab:model}); RSA-4096 uses more cores than
X25519 in both. We therefore write ``at or near saturation'' and describe the regime by what
was observed (next paragraph), not by a threshold.

A throughput deficit is defined by one rule. To say where ML-KEM lowered throughput we apply
one rule to every closed-loop cell in which the Python client ran both ML-KEM and X25519: a
\emph{deficit} if the 90\,\% interval of the paired throughput ratio lies below 0.98, an
\emph{advantage} if it lies above 1.02. This is the rule pre-registered for the native reader;
applying it to the other experiments was decided afterwards and is exploratory
(see the end of this subsection).

The identity becomes a prediction only when applied to data it
was not computed from. We did that once, with a ratio fixed before two follow-up experiments;
it missed (Section~\ref{sec:mechanism}).

There were five deviations from the pre-registration.
(1) The first attempt at \lab{Primary-x86} (batch E) was cut off after 134 of 300 runs when an account-level spending
guard stopped both instances. Its data was never downloaded or analysed; the batch was restarted
from its first run with the same configuration and seed. (2) \lab{Profiling} was run twice,
once under system-call tracing and once under CPU sampling, because the first CPU profiles were
unusable. (3) The stall rule is applied to closed-loop runs only, since open-loop throughput is
fixed by the offered rate. (4) The pre-registration of \lab{Control-Binding} was pushed five minutes after
\lab{Control-Manifest} had started; it concerns \lab{Control-Binding} and a prediction for
\lab{Control-Manifest}, none of whose data had been downloaded. (5) The disk baseline of the
testbed of the follow-up experiments did not complete; its network baseline did.

Several analyses were not pre-specified. The pre-registration fixed the primary family, its test and the
training and held-out split. It did not fix the following, which were chosen during analysis and
are exploratory: applying Holm's correction to the ARM family as well; client CPU per request as a
second endpoint; the CPU accounting described above, the saturation criterion for
the Python client, the sweep of all cells with the deficit rule and the stage comparison of
Section~\ref{sec:mechanism};
the single-thread timing of the native reader's key recovery;
the bootstrap check; the definition of the knee as twice the light-load response time; the
functional form of the load-dependent model; the order-and-time regression of
Section~\ref{sec:validity}; and the layout of the session and hybrid tables.

\section{Results}\label{sec:results}
\subsection{Per-Object Cost}\label{sec:percost}
Table~\ref{tab:handshake} gives key-establishment time per request in the exploratory campaign,
and Table~\ref{tab:ke} in the confirmatory campaign on both architectures. On the x86 client
ML-KEM-768 takes \RevKeMLKEMPut\,ms to encapsulate on PUT and \RevKeMLKEMGet\,ms to decapsulate
on GET, in agreement with the exploratory values (\HsPutPone\,ms and \HsGetPone\,ms). X25519
takes \RevKeEcdhPut\,ms on PUT, where the sender also generates an ephemeral key, and
\RevKeEcdhGet\,ms on GET. RSA-4096 is cheap to encrypt but takes \HsGetRone\,ms to decrypt,
\HsRoneOverPoneGet$\times$ ML-KEM decapsulation. The ARM client is slower in absolute terms
(\ArmKeMLKEMPut\,ms and \ArmKeMLKEMGet\,ms for ML-KEM) with the same ordering of modes. These are
times through a Python binding and include its call overhead; they are not primitive benchmarks.

\begin{table}[t]
\caption{Exploratory campaign: key establishment per request (mean [95\,\% CI], one worker, objects up to 16\,MiB, \lab{Expl-Size}) and manifest size. Manifests were recorded for a 1\,KiB object, except P2, recorded for a 1\,MiB object; the manifest embeds object-dependent fields such as the plaintext size and the object key, so that value differs by 6 bytes from P2's in Table~\ref{tab:meta}, where all modes are recorded for 1\,KiB objects.}
\label{tab:handshake}
\centering\footnotesize
\resizebox{\columnwidth}{!}{
\begin{tabular}{lrrr}
\toprule
Mode & PUT (ms) & GET (ms) & Manifest (B) \\
\midrule
B0 plaintext & -- & -- & 522 \\
B1 AES only & 0.001 [0.001, 0.001] & 0.001 [0.001, 0.001] & 576 \\
R1 RSA-4096 & 0.094 [0.094, 0.094] & 1.486 [1.484, 1.488] & 1{,}276 \\
B2 X25519 & 0.087 [0.086, 0.088] & 0.049 [0.049, 0.049] & 622 \\
P1 ML-KEM-768 & 0.045 [0.045, 0.046] & 0.047 [0.046, 0.047] & 2{,}034 \\
P2 hybrid & -- & -- & 2{,}091 \\
\bottomrule
\end{tabular}
}
\end{table}

\begin{table}[t]
\caption{Confirmatory campaign: per-object key-establishment time (ms, mean of five runs, one worker, 64\,KiB objects; \lab{Hybrid-Matrix} and \lab{OneWorker-ARM}). P2 was not measured on ARM.}
\label{tab:ke}
\centering\footnotesize
\begin{tabular}{lcc|cc}
\toprule
 & \multicolumn{2}{c|}{x86 (ms)} & \multicolumn{2}{c}{ARM (ms)} \\
Mode & PUT & GET & PUT & GET \\
\midrule
B2 X25519 & 0.089 & 0.049 & 0.109 & 0.068 \\
P1 ML-KEM-768 & 0.045 & 0.047 & 0.058 & 0.063 \\
P2 hybrid & 0.133 & 0.097 & -- & -- \\
P3 X-Wing & 0.142 & 0.133 & 0.177 & 0.167 \\
\bottomrule
\end{tabular}

\end{table}

\finding{1}{Per-object ML-KEM-768 costs \RevKeMLKEMPut--\RevKeMLKEMGet\,ms on x86 and
\ArmKeMLKEMPut--\ArmKeMLKEMGet\,ms on ARM in this stack:
less than X25519 on PUT, comparable on GET, and far less than RSA-4096 decryption.}

\subsection{Where the Cost Appears}\label{sec:bottleneck}\label{sec:primary}
\begin{figure}[t]
\centering
\includegraphics[width=\columnwidth]{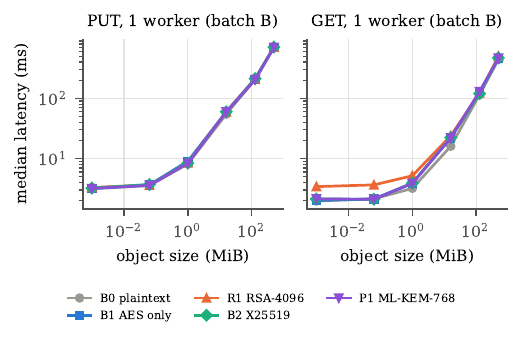}
\caption{Median latency against object size, one worker (\lab{Expl-Size}; log scales).}
\label{fig:size}
\end{figure}

Fig.~\ref{fig:size} shows median latency against object size at one worker. For P1 the share of
request time spent in ML-KEM falls from \KeShareSmallPut\,\% (PUT) and \KeShareSmallGet\,\% (GET)
at 1\,KiB to \KeShareLargePut\,\% and \KeShareLargeGet\,\% at 500\,MiB; even a 1\,KiB PUT takes
\PoneSmallPutMs\,ms, dominated by its two storage round trips. The only mode visibly apart is R1
on GET: its median latency is \HthreeSmallOnePct\,\% above P1 at 64\,KiB and
\HthreeMidOnePct\,\% at 1\,MiB.

\begin{figure*}[t]
\centering
\includegraphics[width=\textwidth]{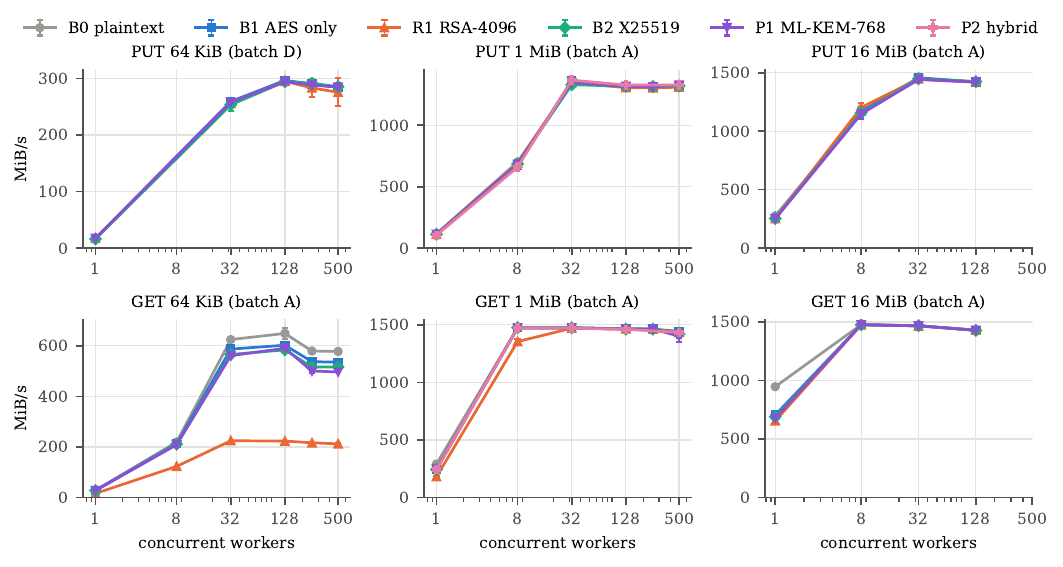}
\caption{Exploratory campaign: throughput (MiB/s, mean and 95\,\% CI over five runs) against concurrent workers. 64\,KiB PUT from \lab{Expl-SmallPUT}, others from \lab{Expl-Sweep}.}
\label{fig:throughput}
\end{figure*}

Fig.~\ref{fig:throughput} shows throughput from 1 to 500 workers. All modes reach the same
ceilings. GETs of 1\,MiB and more saturate the network at \NetCeilingMiB\,MiB/s,
\NetCeilingPct\,\% of the iperf3 baseline; 1\,MiB PUTs level off at about
\PutMidCeilingMiB\,MiB/s. Small PUTs level off at about \PutSmallCeiling{} requests/s while the
client uses at most \PutSmallClientPct\,\% of its CPU, the server \PutSmallServerPct\,\% and the
network \PutSmallNetPct\,\% of the link. Small GETs are limited by the client CPU
(\GetSmallPoneClientPct\,\% for P1 at 128 workers). RSA-4096 reaches that limit much earlier and
delivers \RoneGetThrMin--\RoneGetThrMax{} of P1's throughput at 8--500 workers in the exploratory
campaign and \RevRsaGetLo--\RevRsaGetHi{} at 128 and 500 workers in the confirmatory one.

The small-PUT ceiling is attributed by profiling the server directly. In the exploratory campaign we could only infer
the cause of the small-PUT ceiling by elimination. The confirmatory campaign profiles the server
directly during 1\,KiB and 64\,KiB PUT at 128 workers (Table~\ref{tab:profile}). Three
observations locate it. The device is not the limit: over \RevProfDiskSeconds{} loaded
seconds the data disk's median write latency was \RevProfDiskAwait\,ms (95th percentile
\RevProfDiskAwaitPninetyfive\,ms) with a mean queue depth of \RevProfDiskQueue, while it wrote
\RevProfDiskWriteMib\,MiB/s against an \RevDiskWrite\,MiB/s baseline; requests at that load take
tens of milliseconds. The device's reported utilization was \RevProfDiskUtil\,\%, but for an
NVMe device that figure is the share of time with at least one request in flight, not a measure
of saturation; latency and queue depth are the informative quantities. System-call time goes to locks and the namespace: of the time traced
inside MinIO's system calls, \RevProfFutexTime\,\% is lock waiting (futex) and
\RevProfNamespaceTime\,\% is directory and metadata operations, which are
\RevProfNamespaceCalls\,\% of all calls; writing the data is \RevProfWriteTime\,\%. MinIO does
call \texttt{fdatasync}, but those calls are \RevProfSyncCalls\,\% of all calls and
\RevProfSyncTime\,\% of the traced time. The tracing windows also contain the deletion of
objects after each run, so we report shares and not counts per upload. The CPU profile agrees: \RevProfCpuUser\,\% of MinIO's CPU samples are in
its own code and \RevProfCpuKernel\,\% in the kernel, where lock-spinning routines alone account
for \RevProfCpuKernelLocks\,\% of all samples. (Across the \RevNoisyRuns{} runs of the 64\,KiB PUT
cell at 128 workers, throughput and server CPU use move together, with correlation
\RevNoisyCpuCorr; that is consistent with a server-side origin of the variation but is not
evidence of cause, since both rise with load.) We could not resolve function names inside the MinIO binary, so we attribute
the ceiling to \emph{contention in MinIO's and the file system's metadata path on this host}, not
to a named MinIO component.

\begin{table}[t]
\caption{Where MinIO's system-call time goes during small-object PUT at 128 workers (\RevProfStraceCycles{} tracing windows, \lab{Profiling}). Runs under tracing are slowed and supply no performance numbers.}
\label{tab:profile}
\centering\footnotesize
\begin{tabular}{lrr}
\toprule
System-call group & Traced time & Calls \\
\midrule
Lock waiting (futex) & 44.8\% & 3.4\% \\
Namespace and metadata & 34.8\% & 64.6\% \\
Data write & 5.4\% & 5.1\% \\
Read (network and files) & 5.3\% & 11.0\% \\
Durability (fdatasync) & 1.1\% & 0.5\% \\
\bottomrule
\end{tabular}

\end{table}

Table~\ref{tab:resources} lists CPU and memory for representative workloads of the exploratory
campaign. Per request, P1 uses between \HfourBtwoMin{} and \HfourBtwoMax\,ms more client CPU than
X25519 at one worker, and \HfourRoneGetMin--\HfourRoneGetMax\,ms less than RSA-4096 on GET.
Under contention the gap is larger: at 500 workers on 64\,KiB GET, ML-KEM needs
\RevHotCpuDelta\,\textmu s more than X25519 on x86 and \ArmHotCpuDelta\,\textmu s more on ARM
(Section~\ref{sec:mechanism}). Server CPU does not depend on the mode. Client memory is
dominated by the interpreter and buffers: about \MemPeakSmallSat\,MiB at 500 workers, within \MemModeSpreadPct\,\% across modes;
streaming keeps the growth of a single 500\,MiB PUT at most \MemGrowthLargeOne\,MiB. Server
memory growth has overlapping intervals across modes in \ServerGrowthOverlap{} of
\ServerGrowthWorkloads{} workloads. The one large value in Table~\ref{tab:resources}, P1 on
1\,MiB PUT, is a single run in which MinIO grew its heap by \ServerGrowthOneRun\,MiB; the other
four runs of that cell grew by at most \ServerGrowthOtherRuns\,MiB.

\begin{table*}[t]
\caption{Exploratory campaign: CPU (\% of the host's 16 vCPU, measured window), client CPU time per request and memory (peak RSS and growth over the run), mean of five runs}
\label{tab:resources}
\centering\footnotesize
\begin{tabular}{llrrrrrr}
\toprule
Workload & Mode & Client CPU \% & Server CPU \% & CPU ms/req & Client peak MiB & Client growth MiB & Server growth MiB \\
\midrule
GET 64\,KiB c128 (A) & B1 & 91 & 71 & 1.49 & 2{,}144 & 367 & 4 \\
GET 64\,KiB c128 (A) & R1 & 98 & 23 & 4.36 & 1{,}911 & 125 & 5 \\
GET 64\,KiB c128 (A) & B2 & 92 & 69 & 1.54 & 2{,}134 & 349 & 0 \\
GET 64\,KiB c128 (A) & P1 & 93 & 70 & 1.56 & 2{,}110 & 331 & 1 \\
PUT 1\,MiB c128 (A) & B1 & 34 & 53 & 3.92 & 2{,}283 & 265 & 16 \\
PUT 1\,MiB c128 (A) & R1 & 42 & 51 & 4.97 & 2{,}192 & 116 & 37 \\
PUT 1\,MiB c128 (A) & B2 & 35 & 52 & 4.06 & 2{,}298 & 126 & 33 \\
PUT 1\,MiB c128 (A) & P1 & 35 & 51 & 4.07 & 2{,}291 & 254 & 158 \\
PUT 500\,MiB c16 (B) & B1 & 25 & 47 & 675.42 & 2{,}698 & 174 & 70 \\
PUT 500\,MiB c16 (B) & R1 & 25 & 47 & 676.55 & 2{,}719 & 188 & 27 \\
PUT 500\,MiB c16 (B) & P1 & 26 & 47 & 680.14 & 2{,}729 & 202 & 18 \\
GET 500\,MiB c16 (B) & B1 & 23 & 6 & 1{,}033.11 & 2{,}011 & 12 & 90 \\
GET 500\,MiB c16 (B) & R1 & 23 & 6 & 1{,}024.33 & 2{,}025 & 28 & 144 \\
GET 500\,MiB c16 (B) & P1 & 22 & 6 & 997.48 & 2{,}011 & 33 & 67 \\
\bottomrule
\end{tabular}

\end{table*}

\begin{figure}[t]
\centering
\includegraphics[width=\columnwidth]{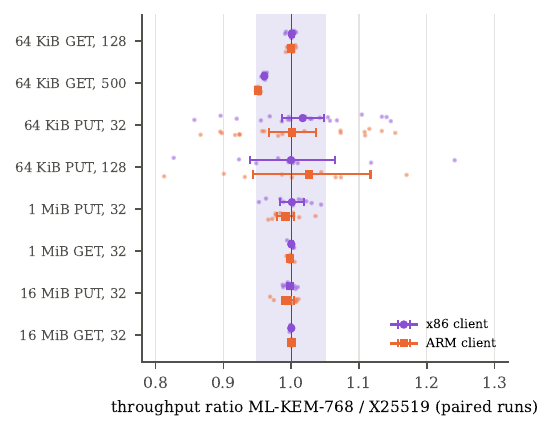}
\caption{Primary family. Small dots: the paired ratio of every repetition. Markers and bars: mean ratio and its 90\,\% interval. Shaded: the $\pm$5\,\% equivalence region.}
\label{fig:tost}
\end{figure}

Table~\ref{tab:tost} and Fig.~\ref{fig:tost} give the pre-registered result
(\lab{Primary-x86}, \lab{Primary-ARM}). Seven of the eight cells were not client-saturated, so in
them no key-establishment mode could have changed throughput; their verdicts confirm where the
bottleneck is and say little about ML-KEM. With that stated: on the x86 client ML-KEM-768 is
equivalent to X25519 in \RevPrimaryEq{} of the 8 cells after Holm correction and inconclusive
in \RevPrimaryInc; no cell is non-equivalent. The ARM replication also yields \ArmPrimaryEq{}
equivalent and \ArmPrimaryInc{} inconclusive cells, though not the same ones. The informative
cell is the saturated one, 64\,KiB GET at 500 workers.

We first look at all cells. Applying the deficit rule of Section~\ref{sec:cpumethod} to all
\AudSweepCells{} closed-loop cells in which the Python client ran both modes, across every
campaign, ML-KEM shows a deficit in \AudSweepDeficits. All of them are
\AudSweepDeficitOps{} of objects of at most \AudSweepDeficitMaxSizeKiB\,KiB with at least
\AudSweepDeficitMinWorkers{} workers, and in all of them the client used at least
\AudSweepDeficitMinCores{} of its 16 cores. Cores in use mark the regime without separating it
sharply: \AudSweepBusyNoDeficit{} other cells were at least that busy and showed no deficit.
ML-KEM shows an advantage in \AudSweepAdvantages{} cells, both \AudSweepAdvantageOps{} at one
worker in the exploratory campaign (Section~\ref{sec:anomalies}).

For large objects the verdict is equivalence throughout. All four cells at 1\,MiB and 16\,MiB are equivalent on both architectures,
with ratios between \RevLargeRatioLo{} and \RevLargeRatioHi{} on x86 and between
\ArmLargeRatioLo{} and \ArmLargeRatioHi{} on ARM. Here the network or the server sets the pace.

The saturated cell is the one that shows a cost. At 64\,KiB GET with 500 workers
ML-KEM delivers \RevHotDeficit\,\% less than X25519 on x86 (90\,\% CI \RevHotCI) and
\ArmHotDeficit\,\% less on ARM (\ArmHotCI). The intervals are narrow, so this is a measured
cost, not noise: on x86 it lies just inside the margin, so the cell is formally equivalent and at
the same time significantly different from 1 (its 95\,\% interval excludes 1), on ARM the
interval touches the bound and the verdict is inconclusive, and in neither case does it pass at
$\pm$2\,\%. At 128 workers the same workload shows no difference.

In small PUT the inconclusive verdicts reflect noise, not a difference. The two 64\,KiB PUT cells have ratios near 1
(\RevNoisyRatio{} and \RevKneeRatio{} on x86) but paired standard deviations of
\RevKneeSd--\ArmNoisySd\,\% in log ratio, the server-side variation described
above. The knee cell passes on its own on x86 ($p=\RevKneeP$) but not after
correction ($p_{\text{Holm}}=\RevKneePHolm$), and passes after correction on ARM. The cell at 128
workers is inconclusive on both: at its observed standard deviation ten repetitions give a single
test a power of only \RevPowerNoisy\,\%, and about \RevPowerNoisyNeeded{} would be needed for
80\,\%. Our power calculation, based on the exploratory campaign's lower variance, was too
optimistic for this cell.

We followed up with adequate power. Because those two cells were under-powered, we
pre-registered and ran them again with 50 repetitions each (\lab{Power-PUT}, \AudKRuns{} runs,
Table~\ref{tab:power}). Both are now equivalent: \AudKKneeRatio{} (90\,\% CI \AudKKneeCI) at 32
workers and \AudKSatRatio{} (\AudKSatCI) at 128, also when the \AudKStalls{} stalled runs are
excluded. This is a new test on new data, on another server instance; it does not change the
verdicts of Table~\ref{tab:tost}, which stand as the outcome of the original family.

\begin{table}[t]
\caption{\lab{Power-PUT}: the two 64\,KiB PUT cells of the primary family repeated with 50 repetitions. Paired throughput ratio ML-KEM-768/X25519, TOST at $\pm$5\,\%, Holm-adjusted over the two cells; the last column gives the number of pairs and the verdict when stalled runs are excluded.}
\label{tab:power}
\centering\footnotesize
\setlength{\tabcolsep}{3pt}
\resizebox{\columnwidth}{!}{
\begin{tabular}{lrccccccl}
\toprule
Cell & $n$ & Ratio & 90\% CI & 95\% CI & SD & $p_{\text{Holm}}$ & Verdict & Without stalls \\
\midrule
64\,KiB PUT, 32 & 50 & 0.994 & 0.984--1.004 & 0.982--1.006 & 4\,\% & $<$0.001 & Eq. & 50: Eq. \\
64\,KiB PUT, 128 & 50 & 0.996 & 0.979--1.013 & 0.976--1.017 & 7\,\% & $<$0.001 & Eq. & 44: Eq. \\
\bottomrule
\end{tabular}
}
\end{table}

The cost is visible in CPU even where throughput does not move.
Table~\ref{tab:cpuprimary} gives the ratio of client CPU per request. Its interval excludes 1 in
\RevCpuPrimaryExcl{} x86 and \ArmCpuPrimaryExcl{} ARM cells. On 64\,KiB GET ML-KEM needs more
CPU than X25519; on 64\,KiB PUT it needs slightly less, as its encapsulation is cheaper than
X25519's ephemeral key generation and exchange.

The verdicts were checked for sensitivity to the test, the margin and stalled runs. A percentile bootstrap reproduces the unadjusted $t$-based verdict in
\RevBootAgree{} of \RevBootTotal{} cells; the exception is the ARM client-bound cell, whose
interval sits on the 0.95 bound and which the bootstrap places just inside it. At $\pm$10\,\% all 8 x86 cells and \ArmPrimaryEqTenPct{} ARM cells are
equivalent; at $\pm$2\,\%, \RevPrimaryEqTwoPct{} and \ArmPrimaryEqTwoPct. Excluding stalled runs
(\RevPrimaryStalledPairs{} pair on x86, \ArmPrimaryStalledPairs{} on ARM) and re-applying the
correction changes \RevPrimaryNoStallFlips{} verdicts on x86 and \ArmPrimaryNoStallFlips{} on ARM:
the ARM knee cell, with one pair fewer, becomes inconclusive
($p_{\text{Holm}}=\ArmKneeNoStallPHolm$). That cell's equivalence on ARM is therefore fragile.

\begin{table*}[t]
\caption{Primary family: paired throughput ratio ML-KEM-768/X25519, TOST with bounds [0.95, 1.05], $\alpha=0.05$, Holm-adjusted over the 8 cells of each client. Eq.\ = equivalent, Inc.\ = inconclusive. The $\pm$2\,\% column reports whether the cell is also equivalent at that stricter margin (unadjusted).}
\label{tab:tost}
\centering\footnotesize
\setlength{\tabcolsep}{4pt}
\begin{tabular}{lrcccclc|cccl}
\toprule
 & \multicolumn{7}{c|}{x86 client (\lab{Primary-x86})} & \multicolumn{4}{c}{ARM client (\lab{Primary-ARM})} \\
Cell (size, op., workers) & $n$ & Ratio & 90\% CI & 95\% CI & $p_{\text{Holm}}$ & Verdict & $\pm$2\% & Ratio & 90\% CI & $p_{\text{Holm}}$ & Verdict \\
\midrule
64\,KiB GET, 128 & 10 & 1.001 & 0.998--1.004 & 0.997--1.004 & $<$0.001 & Eq. & yes & 1.000 & 0.997--1.003 & $<$0.001 & Eq. \\
64\,KiB GET, 500 & 10 & 0.960 & 0.959--0.962 & 0.959--0.962 & $<$0.001 & Eq. & no & 0.951 & 0.950--0.953 & 0.136 & Inc. \\
64\,KiB PUT, 32 & 20 & 1.017 & 0.987--1.048 & 0.980--1.055 & 0.083 & Inc. & no & 1.001 & 0.968--1.036 & 0.040 & Eq. \\
64\,KiB PUT, 128 & 10 & 1.000 & 0.939--1.064 & 0.925--1.080 & 0.092 & Inc. & no & 1.026 & 0.943--1.117 & 0.316 & Inc. \\
1\,MiB PUT, 32 & 10 & 1.001 & 0.984--1.018 & 0.980--1.022 & $<$0.001 & Eq. & yes & 0.992 & 0.979--1.004 & $<$0.001 & Eq. \\
1\,MiB GET, 32 & 10 & 1.000 & 0.998--1.002 & 0.998--1.002 & $<$0.001 & Eq. & yes & 0.998 & 0.996--1.000 & $<$0.001 & Eq. \\
16\,MiB PUT, 32 & 10 & 0.998 & 0.993--1.002 & 0.992--1.003 & $<$0.001 & Eq. & yes & 0.995 & 0.987--1.004 & $<$0.001 & Eq. \\
16\,MiB GET, 32 & 10 & 1.000 & 0.999--1.001 & 0.999--1.001 & $<$0.001 & Eq. & yes & 1.001 & 1.000--1.001 & $<$0.001 & Eq. \\
\bottomrule
\end{tabular}

\end{table*}

\begin{table}[t]
\caption{Primary cells: paired ratio of client CPU per request, ML-KEM-768/X25519, with 95\,\% interval (exploratory second endpoint, unadjusted)}
\label{tab:cpuprimary}
\centering\footnotesize
\begin{tabular}{lcc}
\toprule
Cell & x86 client & ARM client \\
\midrule
64\,KiB GET, 128 & 1.018 [1.015, 1.020] & 1.025 [1.021, 1.028] \\
64\,KiB GET, 500 & 1.046 [1.044, 1.049] & 1.061 [1.060, 1.063] \\
64\,KiB PUT, 32 & 0.990 [0.983, 0.997] & 0.986 [0.982, 0.990] \\
64\,KiB PUT, 128 & 0.992 [0.977, 1.008] & 0.999 [0.970, 1.030] \\
1\,MiB PUT, 32 & 1.002 [0.996, 1.008] & 0.997 [0.995, 0.998] \\
1\,MiB GET, 32 & 1.006 [0.995, 1.018] & 1.005 [0.999, 1.011] \\
16\,MiB PUT, 32 & 1.020 [0.997, 1.043] & 1.031 [1.010, 1.053] \\
16\,MiB GET, 32 & 1.012 [0.994, 1.030] & 1.022 [1.005, 1.040] \\
\bottomrule
\end{tabular}

\end{table}

For context, Table~\ref{tab:h1} summarizes the P1/B2 ratios of the
exploratory campaign over \HoneCells{} cells, judged by its original rule (95\,\% interval inside
$\pm$5\,\%, five runs, no multiplicity control): \HoneEquiv{} equivalent, \HoneInconc{}
inconclusive, \HoneDiff{} different. It showed the same client-bound deficit at 64\,KiB GET and
500 workers (\HoneSatGetPct\,\%). These counts are descriptive and are not a confirmatory result.

\begin{table}[t]
\caption{Exploratory campaign: P1 against X25519 by its original interval rule (equivalent / inconclusive / different, margin $\pm$5\,\%)}
\label{tab:h1}
\centering\footnotesize
\resizebox{\columnwidth}{!}{
\begin{tabular}{llp{1.4cm}p{1.9cm}cc}
\toprule
Batch & Op & Sizes & Workers & Eq./Inc./Diff. & P1/B2 throughput \\
\midrule
A & GET & 1\,MiB, 16\,MiB & 1, 8, 32, 128, 256, 500 & 9/1/0 & 0.981--1.005 \\
A & GET & 64\,KiB & 1, 8, 32, 128, 256, 500 & 6/0/0 & 0.961--1.011 \\
A & PUT & 1\,MiB, 16\,MiB & 1, 8, 32, 128, 256, 500 & 9/1/0 & 0.990--1.014 \\
B & GET & 1\,MiB, 16\,MiB, 128\,MiB, 500\,MiB & 1 & 3/1/0 & 0.921--1.011 \\
B & GET & 1\,KiB, 64\,KiB & 1 & 2/0/0 & 0.991--1.000 \\
B & PUT & 1\,MiB, 16\,MiB, 128\,MiB, 500\,MiB & 1 & 4/0/0 & 1.000--1.039 \\
D & PUT & 1\,KiB, 64\,KiB & 1, 32, 128, 256, 500 & 5/3/0 & 0.989--1.060 \\
\bottomrule
\end{tabular}
}
\end{table}

\finding{2}{Only one of the eight pre-registered cells has a saturated client. In it the Python
client is \RevHotDeficit\,\% (x86) and \ArmHotDeficit\,\% (ARM) slower with ML-KEM than with
X25519. In the other seven the modes are equivalent or, for small-object writes, inconclusive
until repeated with adequate power on x86.}

\subsection{Why the Saturated Client Is Slower}\label{sec:mechanism}\label{sec:native}
The deficit in the saturated cell needs an explanation, because on GET ML-KEM's decapsulation
is \emph{cheaper} than X25519's exchange (Table~\ref{tab:ke}).

Table~\ref{tab:model} applies the identity of
Section~\ref{sec:cpumethod} to 64\,KiB GET. At 500 workers, where the post-quantum modes and
X25519 use the same number of cores to within 0.1, the inverse CPU ratio matches the observed
throughput ratio of ML-KEM and of both hybrids to within \RevModelErrHot{} percentage points on
both architectures, and that of RSA-4096, which needs more than twice the CPU, to within
\RevModelErrHotRsa. At 128 workers the costlier mode uses more cores, throughput holds, and the
identity, which assumes equal cores, overstates the loss by
\RevModelErrMidLo--\RevModelErrMidHi{} points for the post-quantum modes and \RevModelErrMidRsa{}
for RSA-4096. In the saturated cell the deficit of Section~\ref{sec:primary} is therefore a
difference in client CPU: ML-KEM needs \RevHotCpuDelta\,\textmu s more per request than X25519
on x86 (of \RevHotCpuPerRequest\,ms) and \ArmHotCpuDelta\,\textmu s more on ARM.

\begin{table}[t]
\caption{CPU accounting on 64\,KiB GET: the inverse ratio of client CPU per request (column ``Implied by CPU'', computed from the same runs, so not a prediction in the statistical sense) against the observed throughput ratio to X25519 (\lab{Primary-x86} and \lab{Primary-ARM})}
\label{tab:model}
\centering\footnotesize
\setlength{\tabcolsep}{3pt}
\resizebox{\columnwidth}{!}{
\begin{tabular}{lrlccccr}
\toprule
Client & Workers & Mode & Client cores (B2 / mode) & CPU ratio & Implied by CPU & Observed & Error (pp) \\
\midrule
x86 & 128 & P1 ML-KEM-768 & 14.4 / 14.6 & 1.018 & 0.983 & 1.001 & $-$1.8 \\
x86 & 128 & P2 hybrid & 14.4 / 14.7 & 1.050 & 0.953 & 0.977 & $-$2.4 \\
x86 & 128 & P3 X-Wing & 14.4 / 14.8 & 1.080 & 0.926 & 0.955 & $-$2.9 \\
x86 & 128 & R1 RSA-4096 & 14.4 / 15.6 & 2.597 & 0.385 & 0.420 & $-$3.5 \\
x86 & 500 & P1 ML-KEM-768 & 14.8 / 14.9 & 1.046 & 0.956 & 0.960 & $-$0.5 \\
x86 & 500 & P2 hybrid & 14.8 / 14.9 & 1.080 & 0.926 & 0.930 & $-$0.4 \\
x86 & 500 & P3 X-Wing & 14.8 / 14.9 & 1.111 & 0.900 & 0.904 & $-$0.4 \\
x86 & 500 & R1 RSA-4096 & 14.8 / 15.3 & 2.368 & 0.422 & 0.440 & $-$1.7 \\
ARM & 128 & P1 ML-KEM-768 & 13.8 / 14.1 & 1.025 & 0.976 & 1.000 & $-$2.4 \\
ARM & 128 & P3 X-Wing & 13.8 / 14.2 & 1.086 & 0.921 & 0.951 & $-$3.0 \\
ARM & 500 & P1 ML-KEM-768 & 14.2 / 14.4 & 1.061 & 0.942 & 0.951 & $-$0.9 \\
ARM & 500 & P3 X-Wing & 14.2 / 14.3 & 1.125 & 0.889 & 0.896 & $-$0.6 \\
\bottomrule
\end{tabular}
}
\end{table}

The stage timers show where the extra CPU is not. Table~\ref{tab:stages} compares the stage timers of ML-KEM
and X25519 at one worker, where they are not distorted by interpreter-lock waiting. On GET, key
establishment is \emph{faster} with ML-KEM on both architectures. The stage that is slower by
the most is manifest handling, which parses the manifest, decodes the ciphertext and verifies
its hash (+\RevGetStageManifestDelta\,\textmu s on x86, +\ArmGetStageManifestDelta\,\textmu s on
ARM; +\RevPutStageManifestDelta\,\textmu s and +\ArmPutStageManifestDelta\,\textmu s on PUT,
which encodes and hashes it). Key derivation is also slower
(+\RevGetStageKdfDelta\,\textmu s and +\ArmGetStageKdfDelta\,\textmu s on GET) although its
inputs have the same length in both modes (Section~\ref{sec:anomalies}). The cipher is
unaffected.

\begin{table}[t]
\caption{ML-KEM-768 minus X25519 at one worker, 64\,KiB objects (\lab{Hybrid-Matrix} and \lab{OneWorker-ARM}, five runs): difference in mean stage time, and in client CPU per request with its 95\,\% interval}
\label{tab:stages}
\centering\footnotesize
\setlength{\tabcolsep}{4pt}
\begin{tabular}{lrrrrr}
\toprule
 & \multicolumn{4}{c}{Stage time, P1 $-$ B2 (\textmu s)} & (\textmu s, 95\,\% CI) \\
Client, op. & Key est. & Key deriv. & AES-GCM & Manifest & CPU per request \\
\midrule
x86 PUT & $-$44.0 & +4.0 & +0.1 & +16.2 & $-$41 [$-$92, +10] \\
x86 GET & $-$2.1 & +4.6 & +0.2 & +6.4 & +21 [$-$41, +83] \\
ARM PUT & $-$50.6 & +4.8 & +0.1 & +16.4 & $-$22 [$-$42, $-$2] \\
ARM GET & $-$4.7 & +4.6 & +0.2 & +5.5 & +6 [$-$23, +34] \\
\bottomrule
\end{tabular}

\end{table}

The stage timers therefore rule out decapsulation and leave two candidates: the handling of a
manifest \RevMetaExtraMLKEM{} bytes larger, and the way the client reaches ML-KEM. We tested
each with a pre-registered control on 64\,KiB GET at 500 workers
(Table~\ref{tab:control}; decision rules fixed before the runs).

The first control, \lab{Control-Manifest}, tests manifest size. Mode C2 is X25519 whose stored ciphertext is
zero-padded to 1{,}088 bytes, so its manifest is byte-for-byte as long as ML-KEM's. Padding
alone costs \AudPadCost\,\% of throughput (ratio \AudPadRatio, 90\,\% CI \AudPadCI). ML-KEM
remains \AudRestCost\,\% slower than the padded control (\AudRestRatio, \AudRestCI). By the
pre-registered rule, manifest size \emph{\AudPadVerdict} the deficit: it accounts for
\AudPadShare\,\% of it (90\,\% CI \AudPadShareLo--\AudPadShareHi\,\%). At one worker the padded
control shows the same slower manifest stage as ML-KEM (+\AudPadStageGet\,\textmu s against
+\AudMlkemStageGet\,\textmu s), so that stage difference is purely one of size, and it is
small.

The second control, \lab{Control-Binding}, tests the library binding. Our client reaches liboqs through liboqs-python,
which creates a key-encapsulation object for every operation and makes four foreign calls per
decapsulation (allocate, decapsulate, cleanse, free), each releasing the interpreter lock;
X25519, through pyca/cryptography, is a single call. Modes H1 and G1 run the same liboqs
ML-KEM-768 and write the same manifest as P1, but make one foreign call per operation on a
handle created once per process; H1 releases the interpreter lock around it and G1 holds it.
With one call, throughput is equivalent to X25519's within $\pm$2\,\%: H1 reaches
\AudBindReleasedRatio{} (\AudBindReleasedCI) and G1 \AudBindHeldRatio{} (\AudBindHeldCI),
while P1 in the same experiment is at \AudLFullRatio{} (\AudLFullCI). By the
pre-registered rule the binding \emph{\AudBindVerdict} the deficit; it removes
\AudBindReleasedRemoved\,\% of it (90\,\% CI
\AudBindReleasedRemovedLo--\AudBindReleasedRemovedHi\,\%). Whether the single call releases the
lock makes no measurable difference (G1/H1 \AudBindHeldOverReleased, \AudBindHeldOverReleasedCI),
so the cost lies in the per-operation object handling and the number of foreign calls, not in
the locking policy as such. The one-worker timers agree: a decapsulation takes
\AudBindBindingKeOne\,ms through the binding and \AudBindReleasedKeOne\,ms as a single call,
against \AudBindEcdhKeOne\,ms for X25519.

\begin{table*}[t]
\caption{Controls on 64\,KiB GET (\lab{Control-Manifest} and \lab{Control-Binding}, ten repetitions, x86 Python client): paired ratios with 90\,\% intervals. ``Padded manifest'' = X25519 with a manifest as large as ML-KEM's; ``binding'' = ML-KEM-768 through liboqs-python (mode P1); ``one call'' = the same ML-KEM-768 through a single foreign call per operation.}
\label{tab:control}
\centering\footnotesize
\begin{tabular}{llcc}
\toprule
Experiment, workers & Comparison & Throughput ratio & CPU-per-request ratio \\
\midrule
Control-Manifest, 500 & X25519, padded manifest / X25519 & 0.993 [0.991, 0.995] & 1.008 [1.007, 1.009] \\
Control-Manifest, 500 & ML-KEM-768 (binding) / X25519, padded manifest & 0.967 [0.965, 0.969] & 1.038 [1.036, 1.040] \\
Control-Manifest, 500 & ML-KEM-768 (binding) / X25519 & 0.960 [0.958, 0.962] & 1.047 [1.044, 1.049] \\
Control-Manifest, 128 & X25519, padded manifest / X25519 & 0.995 [0.992, 0.998] & 1.005 [1.003, 1.008] \\
Control-Manifest, 128 & ML-KEM-768 (binding) / X25519, padded manifest & 1.004 [1.002, 1.006] & 1.012 [1.010, 1.013] \\
Control-Manifest, 128 & ML-KEM-768 (binding) / X25519 & 0.999 [0.995, 1.002] & 1.017 [1.014, 1.021] \\
Control-Binding, 500 & ML-KEM-768 (binding) / X25519 & 0.963 [0.960, 0.967] & 1.043 [1.040, 1.046] \\
Control-Binding, 500 & ML-KEM-768, one call, lock released / X25519 & 0.999 [0.997, 1.002] & 1.007 [1.004, 1.009] \\
Control-Binding, 500 & ML-KEM-768, one call, lock held / X25519 & 1.002 [0.997, 1.006] & 0.998 [0.995, 1.002] \\
Control-Binding, 500 & ML-KEM-768, one call, lock released / ML-KEM-768 (binding) & 1.038 [1.033, 1.042] & 0.965 [0.963, 0.968] \\
Control-Binding, 500 & ML-KEM-768, one call, lock held / ML-KEM-768, one call, lock released & 1.002 [0.999, 1.006] & 0.991 [0.988, 0.995] \\
Control-Binding, 128 & ML-KEM-768 (binding) / X25519 & 1.005 [1.001, 1.008] & 1.016 [1.013, 1.018] \\
Control-Binding, 128 & ML-KEM-768, one call, lock released / X25519 & 1.000 [0.997, 1.003] & 1.007 [1.004, 1.010] \\
Control-Binding, 128 & ML-KEM-768, one call, lock held / X25519 & 1.002 [0.995, 1.008] & 0.999 [0.994, 1.004] \\
Control-Binding, 128 & ML-KEM-768, one call, lock released / ML-KEM-768 (binding) & 0.995 [0.990, 1.000] & 0.992 [0.987, 0.996] \\
Control-Binding, 128 & ML-KEM-768, one call, lock held / ML-KEM-768, one call, lock released & 1.002 [0.996, 1.008] & 0.992 [0.986, 0.998] \\
\bottomrule
\end{tabular}

\end{table*}

The controls establish the following. On the x86 Python client, where both controls ran, the
deficit in the saturated cell is produced by how the binding invokes the library:
\AudPadShare\,\% of it comes from the larger manifest, and with one foreign call per operation
ML-KEM is equivalent to X25519 within $\pm$2\,\%. The two controls are independent and their
effects need not add up: the one-call modes carry the larger manifest and still match X25519.
The deficit itself was observed in three independent experiments on two server instances
(\RevHotRatio, \AudJFullRatio{} and \AudLFullRatio). The controls were not run on ARM
(Section~\ref{sec:validity}).

The one advance prediction missed. Before \lab{Control-Manifest} and
\lab{Control-Binding} we fixed a predicted ratio of \AudPredicted{} from the CPU ratio measured
in \lab{Primary-x86}. The observed ratios differed from it by \AudPredictionErrorJ{} and
\AudPredictionErrorL{} points, and the prediction lay \AudPredictionInsideL{} both intervals. The
identity holds within an experiment; it did not carry a ratio exactly from one experiment to the
next.

\finding{3}{On the x86 Python client, the \RevHotDeficit\,\% deficit in the saturated cell is
produced by the binding, which makes four foreign calls per decapsulation: with one call
ML-KEM is equivalent to X25519 within $\pm$2\,\% (\AudBindReleasedRatio{} of its throughput).
The larger manifest accounts for \AudPadShare\,\% of the deficit.}

If the Python client's deficit is a binding artifact, does a compiled client show none? We
wrote a reader in Go that implements the GET path of modes B2, P1 and P3 and opens objects
written by the Python client; it does the same per-object work (two S3 reads, manifest parsing,
ciphertext-hash check, key recovery, HKDF, AES-GCM per chunk, a hash of the plaintext), with
ML-KEM-768 and X25519 from Go's standard library. A compiled client needs so little CPU that 16
cores cannot be saturated by this store, so the pre-registered primary cell restricts it to one
core with 128 workers, which saturates it by construction (\lab{Native-GET}, \AudNRuns{} runs,
Table~\ref{tab:native}).

\begin{table*}[t]
\caption{Native Go reader, 64\,KiB GET (\lab{Native-GET}, ten repetitions): X25519 in absolute terms (throughput, process CPU in cores, CPU per request) and paired ratios to it with 90\,\% intervals. Only the one-core cell has a saturated client.}
\label{tab:native}
\centering\footnotesize
\begin{tabular}{lrrr|cc|cc}
\toprule
 & \multicolumn{3}{c|}{X25519} & \multicolumn{2}{c|}{ML-KEM-768 / X25519} & \multicolumn{2}{c}{X-Wing / X25519} \\
Cell & req/s & Cores & CPU (\textmu s) & Throughput & CPU/request & Throughput & CPU/request \\
\midrule
1 core, 128 & 2{,}920 & 1.00 & 343 & 0.905 [0.900, 0.910] & 1.105 [1.099, 1.111] & 0.826 [0.821, 0.831] & 1.211 [1.203, 1.218] \\
16 cores, 128 & 10{,}300 & 9.75 & 947 & 0.943 [0.929, 0.958] & 1.072 [1.051, 1.093] & 0.923 [0.906, 0.940] & 1.117 [1.090, 1.144] \\
16 cores, 500 & 14{,}580 & 11.50 & 789 & 0.996 [0.994, 0.999] & 1.091 [1.054, 1.129] & 0.989 [0.986, 0.992] & 1.157 [1.123, 1.191] \\
\bottomrule
\end{tabular}

\end{table*}

The result for the native reader is as follows. In the saturated cell ML-KEM reaches \AudNOneMlkemThr{} of X25519's throughput
(90\,\% CI \AudNOneMlkemThrCI) and X-Wing \AudNOneXwingThr{} (\AudNOneXwingThrCI): by the
pre-registered rule, a \emph{\AudNOneMlkemVerdict} for both. Their CPU per request is higher by
the inverse factors (\AudNOneMlkemCpu{} and \AudNOneXwingCpu); here the identity of
Section~\ref{sec:cpumethod} holds by construction, since every mode ran on exactly one core, and
this is not a validation of it. On one core the native reader spends
\AudNOneCpuUs\,\textmu s of process CPU per request with X25519.

Part of the native reader's extra CPU can be located. With liboqs called directly, ML-KEM
decapsulation was \emph{cheaper} than X25519 (\AudBindReleasedKeOne\,ms against
\AudBindEcdhKeOne\,ms). In Go's standard library it is dearer: timed alone on the same machine,
decapsulation takes \AudNMicroMlkem\,\textmu s, an X25519 exchange \AudNMicroEcdh\,\textmu s and
an X-Wing decapsulation \AudNMicroXwing\,\textmu s (an exploratory single-thread timing, not
part of the pre-registered experiment). For ML-KEM, key recovery thus accounts for about half of
the extra CPU per request (\AudNMicroMlkemExtra{} of \AudNOneMlkemCpuDelta\,\textmu s); for
X-Wing, for about three quarters (\AudNMicroXwingExtra{} of \AudNOneXwingCpuDelta\,\textmu s).
The remainder was not measured separately in this client.

With 16 cores the reader does not meet the saturation criterion
(\AudNMidCores{} and \AudNHotCores{} cores in use). At 500 workers the modes are within
\AudNHotXwingDeficit\,\% of each other. At 128 workers ML-KEM is \AudNMidMlkemDeficit\,\% and
X-Wing \AudNMidXwingDeficit\,\% slower, with wider intervals (Section~\ref{sec:anomalies}).

\finding{4}{A native reader saturated on one core is \AudNOneMlkemDeficit\,\% slower with ML-KEM
and \AudNOneXwingDeficit\,\% with X-Wing. Across the clients measured, the saturated-client cost
of ML-KEM ranged from equivalence within $\pm$2\,\% (liboqs through one call) to
\AudNOneMlkemDeficit\,\% (Go's standard library).}

\subsection{Hybrid Key Establishment}\label{sec:hybrid}
We first give the cost of the hybrids per object. X-Wing takes \RevKeXwingPut\,ms on PUT and \RevKeXwingGet\,ms on GET on
x86 (Table~\ref{tab:ke}), \RevXwingOverMlkemPut{} and \RevXwingOverMlkemGet{} times ML-KEM
alone. The concatenation hybrid P2 takes \RevKeStudyHybridPut\,ms and
\RevKeStudyHybridGet\,ms; on GET that is close to the sum of its two components
(\RevKeSumGet\,ms), whereas X-Wing takes more. The remainder is the combiner hash and our
Python-level handling of the seed-derived keys, which we did not profile separately. Even so, X-Wing is at most \RevKeShareXwingGet\,\% of a one-worker
64\,KiB request on x86 and \ArmKeShareXwingGet\,\% on ARM, and \RevKeShareXwingLarge\,\% of a
128\,MiB PUT.

Table~\ref{tab:secondary} repeats the primary family for the hybrids
(exploratory, unadjusted). X-Wing is equivalent to X25519 in \RevXwingEq{} of 8 cells on x86 and
\ArmXwingEq{} on ARM. In the client-bound cell (64\,KiB GET, 500 workers) it is non-equivalent on
both: \RevXwingHotDeficit\,\% slower than X25519 on x86 and \ArmXwingHotDeficit\,\% on ARM, and
\RevXwingVsMlkemHotDeficit\,\% and \ArmXwingVsMlkemHotDeficit\,\% slower than ML-KEM alone. At 128
workers the same workload already costs \RevXwingMidDeficit--\ArmXwingMidDeficit\,\%. From
1\,MiB upward X-Wing is equivalent to both on both architectures. P2 is equivalent to P1 in all
\RevStudyHybridEq{} cells, with a lowest ratio of \RevStudyHybridLo.

\begin{table}[t]
\caption{Hybrid modes in the primary cells: paired throughput ratio and unadjusted TOST verdict at $\pm$5\,\% (exploratory). P3 = X-Wing, P2 = concatenation hybrid.}
\label{tab:secondary}
\centering\footnotesize
\setlength{\tabcolsep}{3pt}
\resizebox{\columnwidth}{!}{
\begin{tabular}{lccc|cc}
\toprule
 & \multicolumn{3}{c|}{x86 client} & \multicolumn{2}{c}{ARM client} \\
Cell & P3/B2 & P3/P1 & P2/P1 & P3/B2 & P3/P1 \\
\midrule
64\,KiB GET, 128 & 0.955 (Eq.) & 0.955 (Eq.) & 0.976 (Eq.) & 0.951 (Inc.) & 0.952 (Inc.) \\
64\,KiB GET, 500 & 0.904 (Non-eq.) & 0.942 (Non-eq.) & 0.969 (Eq.) & 0.896 (Non-eq.) & 0.941 (Non-eq.) \\
64\,KiB PUT, 32 & 1.023 (Eq.) & 1.006 (Eq.) & 0.994 (Eq.) & 0.974 (Inc.) & 0.973 (Inc.) \\
64\,KiB PUT, 128 & 0.985 (Inc.) & 0.986 (Inc.) & 1.004 (Eq.) & 1.019 (Inc.) & 0.993 (Eq.) \\
1\,MiB PUT, 32 & 1.009 (Eq.) & 1.008 (Eq.) & 1.005 (Eq.) & 0.981 (Eq.) & 0.989 (Eq.) \\
1\,MiB GET, 32 & 0.998 (Eq.) & 0.998 (Eq.) & 0.999 (Eq.) & 0.998 (Eq.) & 1.000 (Eq.) \\
16\,MiB PUT, 32 & 1.000 (Eq.) & 1.003 (Eq.) & 1.004 (Eq.) & 0.989 (Eq.) & 0.994 (Eq.) \\
16\,MiB GET, 32 & 1.000 (Eq.) & 0.999 (Eq.) & 1.000 (Eq.) & 1.000 (Eq.) & 0.999 (Eq.) \\
\bottomrule
\end{tabular}
}
\end{table}

Table~\ref{tab:hybrid} extends the comparison to \RevMatrixCells{} workloads
from 1\,KiB to 128\,MiB. X-Wing's throughput ratio to X25519 ranges from \RevMatrixXwingLo{} to
\RevMatrixXwingHi{} and lies within $\pm$5\,\% in \RevMatrixXwingWithin{} of them; the lowest
values are again small GETs at 128 workers. Its client CPU per request is up to
\RevMatrixXwingCpuHi\,\% above X25519. Tail latency (p99) follows throughput and shows no
separate hybrid penalty. Client peak memory differs from X25519 by a median of
\RevHybridRssDiffMedian\,MiB (\RevHybridRssDiffMedianPct\,\%) and at most
\RevHybridRssDiffMaxPct\,\% across modes and workloads. One entry of this table, 1\,MiB PUT with
one worker, is not a cryptographic effect (Section~\ref{sec:anomalies}).

\begin{table*}[t]
\caption{\lab{Hybrid-Matrix} (64\,KiB at 128 workers from \lab{Primary-x86}): X25519 in absolute terms and paired ratios of ML-KEM-768 (P1), the concatenation hybrid (P2) and X-Wing (P3) to it. $^{*}$ = the 95\,\% interval of the ratio excludes 1 (unadjusted, five runs; ten in \lab{Primary-x86}). Of the \RevMatrixTests{} ratios, \RevMatrixRawSig{} are marked; \RevMatrixFdrSig{} remain at a false-discovery rate of 5\,\% (Benjamini--Hochberg), so individual marks should not be over-read.}
\label{tab:hybrid}
\centering\footnotesize
\setlength{\tabcolsep}{4pt}
\begin{tabular}{lrrr|ccc|ccc|ccc}
\toprule
 & & \multicolumn{2}{c|}{X25519 (B2)} & \multicolumn{3}{c|}{Throughput ratio} & \multicolumn{3}{c|}{p99 latency ratio} & \multicolumn{3}{c}{Client CPU/request ratio} \\
Workload & $c$ & req/s & MiB/s & P1 & P2 & P3 & P1 & P2 & P3 & P1 & P2 & P3 \\
\midrule
1\,KiB PUT & 1 & 298 & 0.3 & 1.00 & 0.99 & 0.97$^{*}$ & 1.01 & 1.03 & 1.04$^{*}$ & 0.99$^{*}$ & 1.07$^{*}$ & 1.07$^{*}$ \\
1\,KiB GET & 1 & 457 & 0.4 & 0.99 & 0.98 & 1.02$^{*}$ & 1.01 & 1.03$^{*}$ & 1.02$^{*}$ & 1.01$^{*}$ & 1.05$^{*}$ & 1.08$^{*}$ \\
1\,KiB PUT & 128 & 5{,}008 & 4.9 & 1.04 & 1.07 & 1.01 & 0.86 & 0.82 & 0.92 & 1.01 & 1.10$^{*}$ & 1.10$^{*}$ \\
1\,KiB GET & 128 & 11{,}028 & 10.8 & 0.97$^{*}$ & 0.94$^{*}$ & 0.91$^{*}$ & 0.99 & 1.03$^{*}$ & 1.07$^{*}$ & 1.04$^{*}$ & 1.08$^{*}$ & 1.12$^{*}$ \\
64\,KiB PUT & 1 & 250 & 15.6 & 1.01 & 0.99 & 0.99 & 1.01 & 1.03 & 1.03$^{*}$ & 0.97 & 1.03$^{*}$ & 1.04$^{*}$ \\
64\,KiB GET & 1 & 451 & 28.2 & 1.03 & 1.02 & 0.95$^{*}$ & 0.95 & 1.01 & 1.04$^{*}$ & 1.02 & 1.04 & 1.13$^{*}$ \\
64\,KiB PUT & 128 & 4{,}569 & 286 & 1.00 & 1.00 & 0.99 & 0.98 & 0.91 & 0.95 & 0.99 & 1.09$^{*}$ & 1.09$^{*}$ \\
64\,KiB GET & 128 & 9{,}415 & 588 & 1.00 & 0.98$^{*}$ & 0.96$^{*}$ & 0.98$^{*}$ & 1.01$^{*}$ & 1.04$^{*}$ & 1.02$^{*}$ & 1.05$^{*}$ & 1.08$^{*}$ \\
1\,MiB PUT & 1 & 115 & 115 & 1.00 & 0.90$^{*}$ & 0.94$^{*}$ & 1.00 & 1.10$^{*}$ & 1.09$^{*}$ & 1.00 & 1.33$^{*}$ & 1.21$^{*}$ \\
1\,MiB GET & 1 & 247 & 247 & 1.00 & 0.98$^{*}$ & 0.97$^{*}$ & 1.00 & 1.03$^{*}$ & 1.02 & 1.00 & 1.02$^{*}$ & 1.04$^{*}$ \\
1\,MiB PUT & 128 & 1{,}321 & 1{,}321 & 1.00 & 0.99 & 0.99$^{*}$ & 1.00 & 1.01 & 1.02 & 1.01 & 1.27$^{*}$ & 1.07$^{*}$ \\
1\,MiB GET & 128 & 1{,}464 & 1{,}464 & 1.00 & 1.00 & 1.00 & 1.00 & 0.98 & 0.99 & 1.01$^{*}$ & 1.01$^{*}$ & 1.04$^{*}$ \\
16\,MiB PUT & 1 & 16 & 251 & 0.99 & 0.99 & 0.99$^{*}$ & 1.01 & 1.00 & 1.01 & 1.01 & 1.01 & 1.02 \\
16\,MiB GET & 1 & 43 & 685 & 1.02 & 1.02 & 0.97$^{*}$ & 0.94 & 0.98 & 1.00 & 0.98 & 0.98 & 1.06$^{*}$ \\
16\,MiB PUT & 128 & 89 & 1{,}421 & 1.00 & 1.00 & 0.99$^{*}$ & 0.95 & 0.97 & 0.90$^{*}$ & 1.05$^{*}$ & 1.08$^{*}$ & 1.06$^{*}$ \\
16\,MiB GET & 128 & 89 & 1{,}430 & 1.00 & 1.00 & 1.00 & 1.03 & 1.01 & 1.00 & 1.03$^{*}$ & 1.02$^{*}$ & 1.01 \\
128\,MiB PUT & 1 & 5 & 588 & 1.04$^{*}$ & 1.03 & 0.99 & 0.98 & 0.98 & 1.01 & 0.90$^{*}$ & 0.93 & 1.03 \\
128\,MiB GET & 1 & 8 & 1{,}017 & 0.98 & 0.98 & 1.01 & 1.00 & 1.02 & 0.98 & 1.03 & 1.02 & 0.99 \\
\bottomrule
\end{tabular}

\end{table*}

\subsection{Storage Overhead}\label{sec:metadata}
The lasting cost of post-quantum key establishment is storage. Table~\ref{tab:meta} converts the
per-object manifest into fleet totals. An ML-KEM manifest is \RevMetaMLKEM\,bytes,
\RevMetaExtraMLKEM{} more than X25519; X-Wing adds another \RevMetaXwingOverMLKEM. Two parts of
that difference must be kept apart. The \emph{intrinsic} cost of ML-KEM is its ciphertext,
\RevMetaIntrinsic{} bytes more than an X25519 public key. The remaining
\RevMetaEncodingOverhead{} bytes are our encoding: the manifest is JSON and carries the
ciphertext in base64 (\RevMetaBaseSixtyFour{} characters), plus \RevMetaNameBytes{} bytes for a
longer algorithm name. For one billion objects the difference is \RevMetaBillionTB\,TB
(\RevMetaBillionTiB\,TiB) in this encoding and \RevMetaIntrinsicBillionTB\,TB with a binary
one.
Relative to the payload it is negligible from 1\,MiB upward and larger than the payload itself at
1\,KiB. These are logical bytes: replication or erasure coding multiplies them, and because the
manifest is a separate S3 object in our design, it also doubles the number of stored objects and
write requests, which matters under per-request billing and for inventory scans. Carrying the
manifest in the object's own metadata, as the Amazon S3 Encryption Client does by
default~\cite{s3ec}, would avoid the second object. Amazon S3 limits user-defined metadata to
2\,KB per object~\cite{s3metadata}: our JSON manifests for ML-KEM and both hybrids
(\RevMetaMLKEM--\RevMetaStudyHybrid\,bytes) are at or above that limit, whereas a binary
manifest (a 1{,}088-byte ciphertext and a few hundred bytes of context) would fit, provided the
store accepts binary metadata values or the client pays the base64 expansion once more in the
header. Our storage results therefore describe a two-object, JSON-manifest design, which we
chose so that an interrupted upload is never visible and a manifest can be read without the
data; they overstate both the bytes and the request count of a metadata-based design.

\begin{table}[t]
\caption{Manifest size per object and its scaling, for this study's JSON/base64 manifest. Extra bytes are relative to X25519; \RevMetaIntrinsic{} of ML-KEM's \RevMetaExtraMLKEM{} are ciphertext, the rest is encoding.}
\label{tab:meta}
\centering\footnotesize
\setlength{\tabcolsep}{2.5pt}
\resizebox{\columnwidth}{!}{
\begin{tabular}{lrr|rrr|rrr}
\toprule
 & & & \multicolumn{3}{c|}{Extra storage at object count} & \multicolumn{3}{c}{Share of payload} \\
Mode & Manifest (B) & Extra (B) & $10^6$ & $10^8$ & $10^9$ & 1\,KiB & 64\,KiB & 1\,MiB \\
\midrule
B2 X25519 & 622 & -- & -- & -- & -- & -- & -- & -- \\
P1 ML-KEM-768 & 2{,}034 & 1{,}412 & 1.4\,GB & 141.2\,GB & 1.41\,TB & 138\% & 2.15\% & 0.13\% \\
P2 hybrid & 2{,}085 & 1{,}463 & 1.5\,GB & 146.3\,GB & 1.46\,TB & 143\% & 2.23\% & 0.14\% \\
P3 X-Wing & 2{,}074 & 1{,}452 & 1.5\,GB & 145.2\,GB & 1.45\,TB & 142\% & 2.22\% & 0.14\% \\
\bottomrule
\end{tabular}
}
\end{table}

\finding{5}{Against X25519, ML-KEM-768 adds \RevMetaIntrinsic{} bytes of ciphertext per object,
and \RevMetaExtraMLKEM{} in our JSON encoding: irrelevant for large objects, dominant for
kilobyte-sized ones, and \RevMetaIntrinsicBillionTB--\RevMetaBillionTB\,TB per billion objects
before redundancy.}

\subsection{Bounded Sessions}\label{sec:tradeoff}
A session changes what a compromise exposes.
In P1 every object has its own shared secret, and an attacker who obtains one learns one object.
In N1 the unit of compromise is the session root: whoever reads a root from process memory, a
crash dump, a swap file or a log can derive the data keys of every object it protected, at most
$B_{\max}$ objects written within $T_{\max}$ seconds. A cache shared between tenants would extend
that exposure across tenants, so roots must be scoped to one recipient key, as ours are. This
exposure is distinct from, and much smaller than, compromise of the recipient's long-term
decapsulation key, which exposes all objects in every mode. The theoretical $1/B_{\max}$
operations per object holds only when sessions fill before the time limit, so we report what was
observed.

Table~\ref{tab:session} gives the result for 1\,KiB PUT, the workload where key establishment is
the largest share of a request.

\begin{table*}[t]
\caption{Bounded sessions (N1) against per-object ML-KEM (P1), 1\,KiB PUT, \lab{Sessions}. CPU, throughput and p99 are paired ratios to P1 ($^{*}$ = 95\,\% interval excludes 1; five runs). Objects and bytes per root are what one leaked root would expose.}
\label{tab:session}
\centering\footnotesize
\setlength{\tabcolsep}{5pt}
\begin{tabular}{lrrcccccr}
\toprule
Load & $B_{\max}$ & $T_{\max}$ (s) & KEM/object & CPU & Throughput & p99 & Objects/root (med/p95/max) & Max KiB/root \\
\midrule
closed, 32 workers & 1 & 1 & 1.0000 & 1.019$^{*}$ & 1.021 & 0.99 & 1/1/1 & 1 \\
closed, 32 workers & 1 & 10 & 1.0000 & 1.016 & 1.021 & 0.95 & 1/1/1 & 1 \\
closed, 32 workers & 1 & 60 & 1.0000 & 1.017$^{*}$ & 1.020 & 0.97 & 1/1/1 & 1 \\
closed, 32 workers & 10 & 1 & 0.1001 & 0.979$^{*}$ & 1.062 & 0.92 & 10/10/10 & 10 \\
closed, 32 workers & 10 & 10 & 0.1001 & 0.972$^{*}$ & 1.016 & 0.97 & 10/10/10 & 10 \\
closed, 32 workers & 10 & 60 & 0.1001 & 0.972$^{*}$ & 1.010 & 0.98 & 10/10/10 & 10 \\
closed, 32 workers & 100 & 1 & 0.0101 & 0.971$^{*}$ & 1.033 & 0.96 & 100/100/100 & 100 \\
closed, 32 workers & 100 & 10 & 0.0101 & 0.963$^{*}$ & 0.966 & 1.05 & 100/100/100 & 100 \\
closed, 32 workers & 100 & 60 & 0.0101 & 0.970$^{*}$ & 1.043 & 0.95 & 100/100/100 & 100 \\
closed, 32 workers & 1{,}000 & 1 & 0.0038 & 0.969$^{*}$ & 1.015 & 0.98 & 248/276/310 & 310 \\
closed, 32 workers & 1{,}000 & 10 & 0.0011 & 0.970$^{*}$ & 1.028 & 0.98 & 1000/1000/1000 & 1{,}000 \\
closed, 32 workers & 1{,}000 & 60 & 0.0011 & 0.970$^{*}$ & 1.025 & 0.97 & 1000/1000/1000 & 1{,}000 \\
open, 1{,}500/s & 100 & 1 & 0.0110 & 0.972$^{*}$ & 1.003 & 0.98 & 94/100/100 & 100 \\
open, 1{,}500/s & 100 & 10 & 0.0102 & 0.973$^{*}$ & 0.999 & 1.00 & 100/100/100 & 100 \\
open, 1{,}500/s & 100 & 60 & 0.0102 & 0.973$^{*}$ & 1.001 & 1.01 & 100/100/100 & 100 \\
open, 1{,}500/s & 1{,}000 & 1 & 0.0107 & 0.972$^{*}$ & 1.000 & 0.99 & 94/110/125 & 125 \\
open, 1{,}500/s & 1{,}000 & 10 & 0.0013 & 0.971$^{*}$ & 1.003 & 0.97 & 916/980/1000 & 1{,}000 \\
open, 1{,}500/s & 1{,}000 & 60 & 0.0013 & 0.971$^{*}$ & 0.993 & 0.98 & 1000/1000/1000 & 1{,}000 \\
\bottomrule
\end{tabular}

\end{table*}

Operations per object follow $1/B_{\max}$ when sessions fill (\RevSessFullKem{} at
$B_{\max}=1000$, $T_{\max}\ge 10$\,s). When the time limit binds first they do not: at
$T_{\max}=1$\,s a root protected a median of \RevSessShortObjects{} objects under closed-loop
load (\RevSessShortKem{} operations per object) and \RevSessLightObjects{} under light open-loop
load, where even $B_{\max}=100$ was not reached (median \RevSessLightHundredObjects). The time
limit, not the object limit, therefore bounds exposure for lightly loaded clients.

Client CPU per request falls by \RevSessCpuSavingLo--\RevSessCpuSavingHi\,\% for
$B_{\max}\ge 10$, and nearly all of that is reached at $B_{\max}=10$. A session of one object
costs \RevSessBoneCpu\,\% \emph{more} CPU than P1, the bookkeeping overhead of the mechanism.

Throughput and tail latency do not change: in none of the \RevSessConfigs{} configurations
with $B_{\max}\ge 10$ does the throughput interval exclude 1
(\RevSessThroughputLo--\RevSessThroughputHi), because this workload is limited by the server. On a
client-bound workload, 64\,KiB GET at 32 workers, a receiver that caches roots gains
\RevSessGetGain\,\% throughput (95\,\% CI \RevSessGetGainLo--\RevSessGetGainHi\,\%) at
$B_{\max}=100$. This is an upper bound: the run reads 32 objects repeatedly, so every root stays
in the receiver's cache and almost no decapsulation remains; a workload with a large working set
would gain less. The exploratory campaign agrees: N1 saved
\NoneSaveOneMin--\NoneSaveOneMax\,\textmu s of CPU per request at one worker and raised
throughput only for small GETs at 32--128 workers (\NoneGetGainMin--\NoneGetGainMax\,\%). Its
queue-aware variant lowered the bound under light load as designed, re-keying
\QaKemRatioLow$\times$ as often as a fixed bound of 100 at \QaRateLow{} requests/s with no
measurable change in tail latency; we did not show that it is better than a smaller fixed bound.

The measurements support a conservative rule. Per-object ML-KEM is the default. Amortization is
enabled only for a workload in which profiling shows the client CPU to be the bottleneck, and
then with the smallest bound that achieves the gain: in our data $B_{\max}=10$ already removes
most of the CPU cost, so a larger bound buys exposure without benefit. $T_{\max}$ should be set
from the tolerated exposure of an idle client, since it is the limit that binds under light load.
The same reasoning applies to the choice between ML-KEM and a hybrid: where the client is not the
bottleneck the hybrid costs only \RevMetaXwingOverMLKEM{} bytes per object, and where it is, the
cost is the throughput difference reported in Section~\ref{sec:hybrid}.

\finding{6}{Sessions trade compromise scope for a
\RevSessCpuSavingLo--\RevSessCpuSavingHi\,\% saving in client CPU. That saving
becomes throughput only on client-bound workloads; a bound of 10 objects captures most of it.}

\subsection{Open Anomalies}\label{sec:anomalies}
Five observations are not explained by anything above. We list them together so that they are
not read past.

(A1) Key derivation on GET at one worker is slower with ML-KEM. At one worker the key-derivation stage is
+\RevGetStageKdfDelta\,\textmu s (x86) and +\ArmGetStageKdfDelta\,\textmu s (ARM) slower on GET
with ML-KEM than with X25519, although its inputs have the same length in both modes. The
padded-manifest control does not show it (+\AudKdfOverPadGet\,\textmu s for ML-KEM over the
padded control), so it is not an effect of manifest size. We have no explanation.

(A2) The native reader shows a deficit without saturation. In \lab{Native-GET} with 16 cores and 128
workers ML-KEM is \AudNMidMlkemDeficit\,\% and X-Wing \AudNMidXwingDeficit\,\% slower than
X25519, in a cell that does not meet the pre-registered saturation criterion
(\AudNMidCores{} cores in use). Two cautions apply: the 16 virtual CPUs are 8 physical cores, and
the reader's CPU per request is two to three times higher with 16 cores than with one, which
suggests runtime overhead that makes ``cores in use'' a soft measure of saturation here. This
result is the one exception to the statement that ML-KEM lowered throughput only at or near
client saturation.

(A3) X-Wing is below X25519 in a Python cell we call unsaturated. At 64\,KiB GET with 128
workers X-Wing is \RevXwingMidDeficit\,\% (x86) and \ArmXwingMidDeficit\,\% (ARM) slower than
X25519 (Table~\ref{tab:secondary}), although by our criterion that cell is not saturated. The
client is the busiest resource there, so this may be near-saturation; we did not test it.

(A4) In some exploratory cells ML-KEM was faster than X25519 at one worker. By the deficit rule, ML-KEM has an
advantage in \AudSweepAdvantages{} cells of the exploratory campaign, both PUT at one worker
(ratios \AudSweepAdvantageLo{} and \AudSweepAdvantageHi, five runs each). Key establishment is
far too small a share of those requests to produce a difference of that size, and the
corresponding cells of \lab{Hybrid-Matrix} do not show an advantage by the same rule. We have
no explanation; an allocator effect of the kind described next is a candidate that we did not
test.

(A5) An allocator effect appears at 1\,MiB. In \lab{Hybrid-Matrix}, at 1\,MiB PUT with one worker X-Wing is
\RevMidPutXwingDeficit\,\% and P2 \RevMidPutStudyHybridDeficit\,\% slower than X25519 although key
establishment is \RevKeShareXwingMidPut\,\% of the request; the difference sits in the AES-GCM
stage timer, which all modes share. A diagnostic pair of experiments (\lab{Allocator}, \AudAllocRuns{} runs,
Table~\ref{tab:alloc}) points to the memory allocator. With default settings the AES-GCM stage
takes between \AudAllocDefaultLo{} and \AudAllocDefaultHi\,ms depending on the mode, a factor of
\AudAllocDefaultSpread. With glibc's dynamic mmap threshold disabled it takes
\AudAllocFixedLo--\AudAllocFixedHi\,ms in every mode, a factor of \AudAllocFixedSpread, and the
throughput spread between modes falls from \AudAllocDefaultThrSpread\,\% to
\AudAllocFixedThrSpread\,\%. The stage time is thus governed by where the 1\,MiB chunk buffer is
allocated, which depends on a mode's earlier allocations, and not by the cipher or the key
establishment. A small residual for X-Wing remains.
The residual for X-Wing is not explained, and the diagnostic was exploratory.

\begin{table}[t]
\caption{1\,MiB PUT, one worker: AES-GCM stage time and throughput with the default allocator and with glibc's mmap threshold fixed at 256\,MiB (\lab{Allocator}, five runs each)}
\label{tab:alloc}
\centering\footnotesize
\begin{tabular}{lrr|rr}
\toprule
 & \multicolumn{2}{c|}{Default allocator} & \multicolumn{2}{c}{Fixed threshold} \\
Mode & AES-GCM (ms) & req/s & AES-GCM (ms) & req/s \\
\midrule
B2 X25519 & 0.12 & 115 & 0.29 & 109 \\
P1 ML-KEM-768 & 0.12 & 116 & 0.29 & 109 \\
P2 hybrid & 0.38 & 104 & 0.29 & 108 \\
P3 X-Wing & 0.28 & 108 & 0.32 & 105 \\
\bottomrule
\end{tabular}

\end{table}

\section{Threats to Validity}\label{sec:validity}\label{sec:pending}
We first examine run-to-run variation.
Fig.~\ref{fig:tost} shows every paired ratio of the primary family. Six cells are tight: the
coefficient of variation of throughput across runs is at most \RevCvQuietMax\,\%. The two
64\,KiB PUT cells vary by \RevCvNoisyLo--\RevCvNoisyHi\,\% per mode. That variation originates at
the server (Section~\ref{sec:bottleneck}) and affects all modes alike, which is why it widens
intervals without shifting ratios.

The pre-registered stall rule flags \RevStalls{} of \RevRuns{} x86 runs and \ArmStalls{} of \ArmRuns{}
ARM runs, all of them PUTs of at most 64\,KiB. Excluding them leaves the 8 x86 verdicts
unchanged and turns one ARM verdict, the knee cell, from equivalent to inconclusive
(Section~\ref{sec:primary}).

Order and time effects were checked as follows. Modes are shuffled within each repetition. As a check, for each primary cell we regress log
throughput on the mode, repetition fixed effects, elapsed time in the experiment, position in the
block and whether the previous run was a PUT. The adjusted P1/B2 ratio differs from the paired
ratio by at most \RevOrderShiftMax\,\% in any cell on either architecture.

The ARM replication reproduces the pattern: equivalence at 1\,MiB and above, a deficit of
\RevHotDeficit\,\% (x86) and \ArmHotDeficit\,\% (ARM) in the client-bound cell, and noise in small PUT. Because that testbed also has another server
instance, it tests the robustness of the ratios and not ARM against x86 in absolute terms.

Threats to internal validity are the following. Stage timers above one worker include interpreter-lock waiting, so
key-establishment time is reported at one worker only. During the x86 chain of the confirmatory
campaign, completed experiments were downloaded from the client while parts of
\lab{Hybrid-Matrix}, \lab{Sessions} and \lab{Open-Loop} were running; the transfer used client CPU and network and may have added noise to those runs.
Mode order is randomized within repetitions, so it cannot favour a mode, and no transfer
overlapped \lab{Primary-x86} or \lab{Primary-ARM}. Modes differ in manifest size as well as
in key establishment; \lab{Control-Manifest} separates the two for the Python client only
(Section~\ref{sec:mechanism}). Seven of the eight primary cells could not have shown an effect
of the mode on throughput. The exploratory campaign's pause failure
(Section~\ref{sec:method}) added noise to small PUT cells; its results for those cells come from
\lab{Expl-SmallPUT}. The 1\,MiB one-worker slowdown of the hybrids is attributed to the allocator by an exploratory
diagnostic and keeps an unexplained residual (Section~\ref{sec:anomalies}). The profiling
attribution names a subsystem, not a function. The two inconclusive primary cells were
under-powered; \lab{Power-PUT} repeated them on x86 only, so on ARM 64\,KiB PUT at 128 workers
and 64\,KiB GET at 500 workers remain inconclusive. Both controls,
\lab{Control-Manifest} and \lab{Control-Binding}, ran on the x86 client only: the ARM client
uses the same binding and shows a similar deficit, but the attribution to the binding is not
tested there. The line we draw between a saturated and an unsaturated Python cell rests on a
difference of a few tenths of a core, was drawn after the data and was not pre-registered;
anomaly A3 may be a consequence of drawing it too high.

Threats to external validity are the limits of what the data supports. Most of the study uses one
Python client with liboqs through a binding, and \lab{Control-Binding} shows how much that instrument can
shape a result: its one client-bound deficit was produced by the binding. The native reader
removes the interpreter but covers only the GET path of three modes at one object size, uses
different cryptographic implementations, and reaches saturation only when confined to one core,
which is an artificial condition; a production client saturating many cores may behave
differently. The baseline set is narrow: no ML-KEM-1024, no RSA-3072, and no remote
key-service wrap, which is a common deployment path. MinIO ran as a single node without erasure
coding or replication, in a community edition that is no longer developed; a distributed
deployment has a different write path and may have a different ceiling and knee. Both hosts sat
in one placement group with a sub-millisecond round trip; a wide-area client would be dominated
by the network. All runs used AWS, in two regions, with one server instance type. liboqs is
intended for prototyping, not production.

Threats to construct validity concern what the measures mean. Throughput equivalence within $\pm$5\,\% is an engineering criterion, not a
statement that two modes are identical; a ratio of \RevHotRatio{} with a narrow interval is
formally equivalent and still a real difference. The CPU accounting is an identity on the runs
it describes; in the native reader it holds by construction; its one advance prediction missed
by up to \AudPredictionErrorL{} points and lay outside both intervals. The statement that a
throughput cost appears only at or near client saturation is contradicted by anomaly A2 if
``saturation'' is read strictly, and the Python client's process CPU and the native reader's
are not measured the same way (thread CPU at 500 workers against process CPU on one core), so
their per-request figures should not be compared as like for like. The exploratory verdict counts are not corrected for
multiplicity. Runs last 20\,s; slower effects such as compaction or long-term memory growth are
not observed.

The following have not been run, and the paper's claims are limited accordingly.
(1)~\emph{A full native client:} PUT, the remaining modes and several object sizes, saturating
many cores, with a second ML-KEM implementation so that library and language can be separated.
(2)~\emph{A manifest control in the native reader}, to split its ML-KEM cost into key recovery
and manifest handling by measurement and not by a separate timing.
(3)~\emph{An HPKE-conformant envelope}, to confirm that the per-object costs transfer.
(4)~\emph{Wider baselines:} ML-KEM-1024, RSA-3072 and a remote key-service wrap.
(5)~\emph{A distributed or second S3 implementation}, and runs of 10 to 30 minutes, for the
store-side results.
(6)~\emph{The two controls on the ARM client.}
(7)~\emph{Multi-recipient objects}, which would need a key-committing construction.

\section{Discussion and Conclusion}\label{sec:discussion}\label{sec:conclusion}
A practitioner learns more from these measurements than the back-of-envelope estimate gives. That tens of microseconds of KEM are
small against milliseconds of storage is true but incomplete. Three things could not have been
read off that estimate. First, the regime in which a cost appeared is narrow and identifiable:
a client at or near CPU saturation, with the size of the cost given by client CPU per request.
Second, the first cause we suspected was wrong. The 1{,}088-byte ciphertext and its handling
explain \AudPadShare\,\% of the x86 Python client's deficit; the rest came from a binding that
crosses into native code four times per decapsulation. Third, the size of the effect depends on
the library: the ML-KEM operation was cheaper than an X25519 exchange with liboqs called
directly and dearer in Go's standard library, and the throughput cost under saturation ranged
accordingly from equivalence within $\pm$2\,\% to \AudNOneMlkemDeficit\,\%. In no saturated cell
was ML-KEM faster than X25519.

For migration planning, the CPU accounting of Section~\ref{sec:cpumethod} gives a short
procedure. Measure client CPU per request for the current and the candidate mode on the real
workload and client. If clients are well below CPU saturation, our data gives no reason to
expect a change in throughput, with the reservation of anomaly A2. If they are saturated, expect
throughput to change by about the inverse ratio of the two CPU figures:
\RevHotDeficit--\ArmHotDeficit\,\% in our Python client, \AudNOneMlkemDeficit\,\% in our native
reader on one core, and equivalence within $\pm$2\,\% with a one-call binding on x86. A
measurement on one client stack does not transfer to another.

For library authors the lesson concerns bindings. A binding that allocates an object and makes several foreign calls
per operation can cost more under thread contention than the cryptography it wraps. In our x86
client, holding a handle per process and making one call per operation removed
\AudBindReleasedRemoved\,\% of the deficit (90\,\% CI
\AudBindReleasedRemovedLo--\AudBindReleasedRemovedHi\,\%); whether that call released the
interpreter lock made no measurable difference.

The cost that does not depend on the implementation is ML-KEM's
1{,}088-byte ciphertext: \RevMetaIntrinsic{} bytes more per object than an X25519 key, or
\RevMetaExtraMLKEM{} more in a JSON/base64 manifest, which puts such a manifest at or above the
2\,KB that Amazon S3 allows for user-defined metadata (Section~\ref{sec:metadata}). For hybrids
there is in addition a client-CPU increment under saturation. RSA-4096 is the poor choice: its
private-key operation makes small-object reads CPU-bound at \RevRsaGetLo--\RevRsaGetHi{} of
ML-KEM's throughput in the Python client.

On the capacity of the store, the open-loop measurements are reported in the supplementary
material because they describe the store and not post-quantum cryptography. Their result is that
mean response time doubles at \RevQMidRhoKneeLo--\RevQSmallRhoKneeHi{} of peak closed-loop
throughput, in line with the known gap between closed- and open-loop
behaviour~\cite{schroeder2006}; the exploratory campaign found that knee in the same place for
the symmetric-only mode and for ML-KEM.

In conclusion, we evaluated per-object post-quantum key establishment on an S3-compatible store with a Python
client on x86 and ARM and with a native Go reader, against a single-node MinIO server on AWS. In
these configurations ML-KEM lowered throughput against X25519 only where the client was at or
near CPU saturation, with one unexplained exception in the native reader
(Section~\ref{sec:anomalies}). In the one saturated cell of the pre-registered family the
Python client was \RevHotDeficit\,\% (x86) and \ArmHotDeficit\,\% (ARM) slower; two
pre-registered controls on x86 attribute that deficit to the library binding and
\AudPadShare\,\% of it to the larger manifest, and with one foreign call per operation ML-KEM was
equivalent to X25519 within $\pm$2\,\%. A native reader saturated on one core was
\AudNOneMlkemDeficit\,\% slower with ML-KEM and \AudNOneXwingDeficit\,\% with X-Wing; a slower
key recovery in its library accounts for about half of the extra CPU for ML-KEM and about three
quarters for X-Wing. In the saturated cells the throughput loss matched the extra client CPU per
request, which makes CPU per request the quantity to measure before a migration. The size of
the cost depends on the client implementation. The cost that every implementation pays is
storage: \RevMetaIntrinsic{} bytes more per object than X25519.

These results describe two clients, one single-node store and one cloud provider, and the
native reader covers the GET path only. They should not be extended to distributed object
stores, other S3 implementations, other libraries or wide-area deployments without new
measurements (Section~\ref{sec:pending}).

\section*{Data and Code Availability}
Code, configurations, the pre-registration, per-run summaries, the raw-data validator, the
native reader (\texttt{native/}) and the analysis scripts that generate every figure, table and
number in this paper (\texttt{analysis/summarize.py}, \texttt{final\_analysis.py},
\texttt{make\_paper\_assets.py}, \texttt{revision\_assets.py}, \texttt{audit\_assets.py}) are in
the project repository, organized following ACM artifact guidelines~\cite{acmbadges}; the
configuration files record the random seed of every experiment, and the repository holds an
export of the hosting service's push log. The repository is public at
\url{https://github.com/revanthreddy0906/pqc-object-store}; it has no persistent identifier. The raw
request records (about 11\,GB compressed) are not in the repository because of their size and are
available from the corresponding author on request.

\bibliographystyle{IEEEtran}
\bibliography{references}

\end{document}